\documentclass[10pt]{article}

\usepackage{graphicx}
\usepackage{float}
\usepackage{algorithm}
\usepackage{bm}
\usepackage{color}
\usepackage{appendix}
\usepackage{subfig}
\usepackage{float}
\usepackage{xpatch}
\usepackage{lipsum}

\usepackage[utf8]{inputenc} 
\usepackage[T1]{fontenc}    
\usepackage[labelfont=sc]{caption}
\usepackage{amsmath,amsbsy,amssymb}
\usepackage{hyperref}
\usepackage{authblk}
\usepackage{enumerate}

\usepackage[]{geometry}
\providecommand{\definitionname}{Definition}
\providecommand{\propositionname}{Proposition}
\providecommand{\theoremname}{Theorem}

\newtheorem{thm}{\protect\theoremname}

\DeclareMathOperator{\sech}{sech}

\newcommand\blfootnote[1]{%
  \begingroup
  \renewcommand\thefootnote{}\footnote{#1}%
  \addtocounter{footnote}{-1}%
  \endgroup
}

\let\citep\cite
\let\citet\cite

\makeatletter
\def\@fnsymbol#1{\ensuremath{\ifcase#1\or \dagger\or \ddagger\or
   \mathsection\or \mathparagraph\or \|\or **\or \dagger\dagger
   \or \ddagger\ddagger \else\@ctrerr\fi}}
\makeatother

\begin{document}

\title{Objective early identification of kinematic instabilities in shear flows}

\author[1]{Bjoern F. Klose*}
\author[2,3]{Mattia Serra*}
\author[1]{Gustaaf B. Jacobs%
 \thanks{Corresponding author. Email: gjacobs@sdsu.edu}}

\affil[1]{Department of Aerospace Engineering, San Diego State University, San Diego, CA 92182, USA}
\affil[2]{School of Engineering and Applied Sciences, Harvard University, Cambridge, MA 02138, USA}
\affil[3]{University of California San Diego, Department of Physics, CA 92093, USA}

\maketitle

\begin{abstract}
A kinematic approach for the identification of flow instabilities is proposed. 
By defining a flow instability in the Lagrangian frame as the increased folding of lines of fluid particles, subtle perturbations and unstable growth thereof are detected early based solely on the curvature change of material lines over finite time.
The material line curvature is objective, parametrization independent, and can be applied to flows of general complexity without knowledge of the base flow. 
An analytic connection between the growth of Eulerian velocity modes perturbing a general shear flow and the induced flow map and Lagrangian curvature change is derived.
The approach is verified to capture instabilities promptly in a temporally developing jet flow, an unstable separated shear flow over a cambered airfoil, and in the onset of a wake instability behind a circular cylinder. 
\end{abstract}

\section{Introduction} \label{introduction}
The path to turbulent flow is paved by linear and non-linear instabilities.
Central to the characterization of  instabilities is whether a local disturbance to an otherwise unperturbed base flow grows or recedes in space and time.
In linear stability analysis (LSA), for example,  a disturbance is assumed to have the form of complex waves. The stability of these waves can be determined by an eigenvalue problem that follows from substitution into linearized governing flow equations \citep{HM90}. \blfootnote{* B.F.K. and M.S. contributed equally}
The theoretical treatment of  instabilities  has been subject of research for many decades (see \citet{BOT88}, \citet{HM90}, \citet{drazin_2003}, \citet{Schmid07}, and \citet{Theofilis11}, for a comprehensive overview).   

Many flows, however, do not have a well-defined base flow. Either, there are  multiple possible frame of references and/or a complex time dependency makes it impossible to  define a base flow, preventing LSA.
As stated in (\citet{drazin_2003}, p. 354) 
``the meaning of instability is not clear when the magnitude of the basic
flow changes substantially with time'' and 
``the method of normal modes is not applicable''.
Simple examples include the flow over a moving flat plate initially at rest \citep{drazin_2003} or the onset of vortex shedding in the wake behind a circular cylinder \citep{SS90}. 
More complex examples are plentiful, such as the flow interaction of two flapping wings or maneuvering flying objects (see for example the flight mechanics of a dragonfly \citep{BXW07}).

Moreover, applying LSA to flows of general complexity can be challenging.
While  methods have been proposed  to this end (\citet{Theofilis11,RUG20}), they are recent and require significant computational resources
and complicated algorithms whose limitations and by extension generality remain to be assessed. 
Even more importantly, perhaps, is that the stability properties of flows often vary in time and non-normal growth can drive the perturbations rather than the most unstable mode predicted by LSA \citep{Schmid07}. 
The introduction of a finite-time horizon to the stability analysis and description of flow instabilities with a non-modal approach has been shown to reveal new perturbations dynamics \citep{SH02,Schmid07}.

Since there is no limit to the complexity of flow problems, an instability should not depend on the frame of the observer. 
An analysis based on velocity components or the streamlines is inherently non-objective,
as these are not invariant under coordinate transformations and always defined with respect to a reference frame.
Velocity gradient based flow identifiers and visualizers, including vorticity and other Eulerian vortex identification criteria, such as the Q-criterion by \citet{HWM88} or the $\lambda_2$-criterion by \citet{JH95}, are frame-invariant in the Galilean sense, i.e. their values are consistent in coordinate systems moving at a constant relative speed. 
A quantity that remains invariant under \textit{any} rotation or translation of the form 
\begin{equation}
\label{eq:obj}
\mathbf{y} = \mathbf{Q}(t)\mathbf{x} + \mathbf{b}(t),
\end{equation}
is called \textit{objective}, where $\mathbf{Q}(t)$ is an orthogonal tensor and $\mathbf{b}(t)$ denotes a translation vector.
From this argument,  \citet{Haller14,SerraHaller2015,HHFH16} propose objective definitions of a vortex.

Material invariance is also the basis for the objective definition of hyperbolic Lagrangian Coherent Structures (LCS) \citep{Haller14} and Objective Eulerian Coherent Structures (OECSs) \citep{SerraHaller2015}.
In general, hyperbolic LCS are manifolds that attract or repel fluid material and can be identified through ridges in the Finite-Time Lyapunov Exponent (FTLE) fields \citep{haller00}. 
Although maxima in the FTLE identify potential LCS, it has been shown that non-hyperbolic ridges can occur in regions of high shear \citep{Haller02}. 
\citet{Haller11} presents a methodology to resolve this inconsistency, but admits that the required computation is challenging and sensitive to noise. 
\citet{BFHS17} discuss the computation of reduced-order FTLEs based on the optimally time-dependent (OTD) modes associated with finite-time instabilities.
The OTD modes were previously introduced by \citet{BS16} and are shown to capture transient instabilities and their direction.
For a more comprehensive literature overview of existing methods to identify transient instabilities and perform dimensionality reduction, we refer the reader to \citet{BFHS17} and the references therein.

Invariably, both the FTLE and the variational Coherent Structures \citep{Haller14,SerraHaller2015} are based solely on the stretching of fluid material. Divergence of particle trajectories, however, may require a long integration time to manifest. This is an undesired feature if one seeks to detect unstable modes in shear flows early.
Only recently, \citet{serra18,serra20,KSJ20a} have focused on more subtle short-term events in Lagrangian fluid trajectories. 
Using the curvature of near-wall material lines, they identified the point early fluid upwelling from a no-slip boundary, and showed that such locations remain hidden to existing techniques. 
The material line curvature is independent of its parametrization, invariant to frame changes under \eqref{eq:obj} and naturally combines stretching- and rotation-based quantities \citep{serra18}.
The objective identification
of early and subtle changes in material lines naturally fits to the identification of instabilities, but a definition
of instability based on this perspective has not been coined yet.

\begin{figure}
\centering
    \begin{minipage}{0.5\textwidth}
    \centering
    \includegraphics[width=0.8\textwidth]{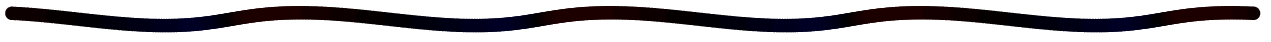}
    \makebox[\textwidth][c]{(a) $t$ = $t_0$}
    \vskip\baselineskip
    \vskip\baselineskip
    \includegraphics[width=0.8\textwidth]{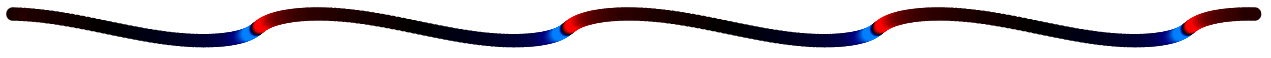}
    \makebox[\textwidth][c]{(b) $t$ = $t_1$}
    \vskip\baselineskip
    \vskip\baselineskip
    \includegraphics[width=0.8\textwidth]{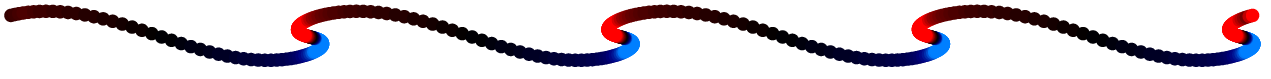}
    \makebox[\textwidth][c]{(c) $t$ = $t_2$}
    \vskip\baselineskip
    \vskip\baselineskip
    \includegraphics[width=0.8\textwidth]{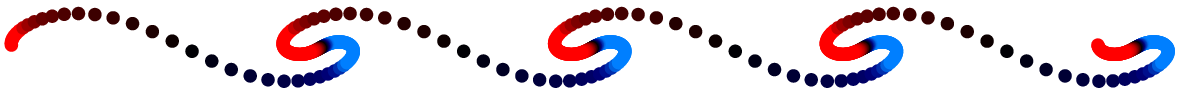}
    \makebox[\textwidth][c]{(d) $t$ = $t_3$}
    \end{minipage}%
    \begin{minipage}{0.5\textwidth}
    \centering
    \includegraphics[width=0.8\textwidth]{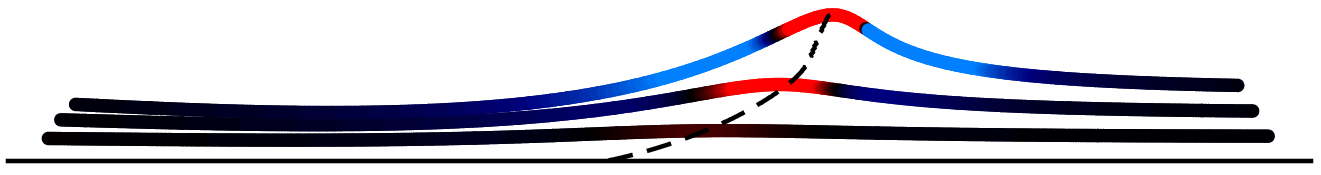}
    \makebox[\textwidth][c]{(e) $t$ = $t_0$}
    \vskip\baselineskip
    \includegraphics[width=0.8\textwidth]{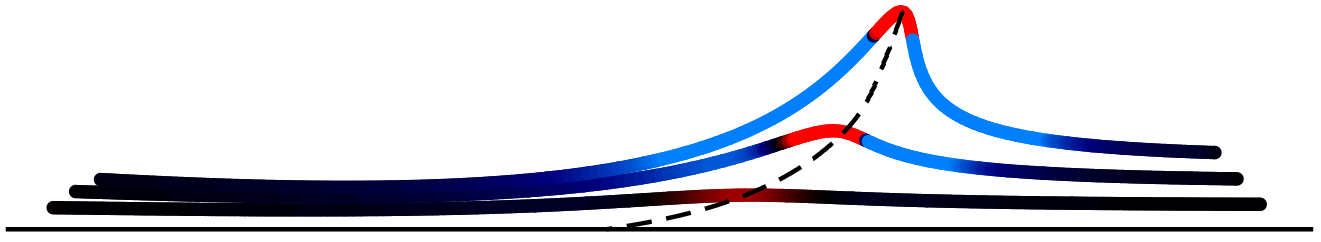}
    \makebox[\textwidth][c]{(f) $t$ = $t_1$}
    \vskip\baselineskip
    \includegraphics[width=0.8\textwidth]{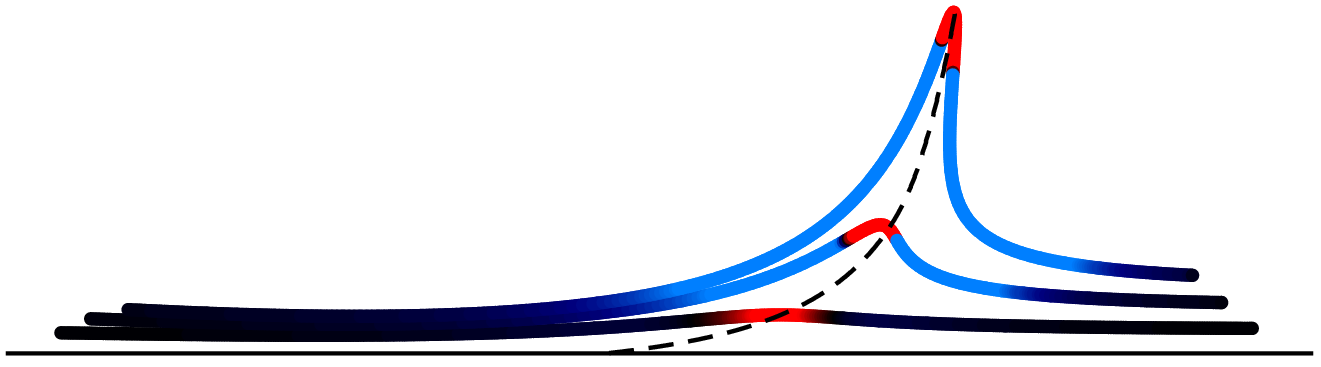}
    \makebox[\textwidth][c]{(g) $t$ = $t_2$}
    \end{minipage}
    \caption{ Temporal development of a  material line in an unstable shear layer (a -- d).
    Time development of  material lines and backbone of separation (dashed) in a separated flow in the
    vicinity of a no-slip wall (e -- g) . 
    Coloring by material line curvature from blue (negative) to red (positive).}
	\label{fig:material_line}
\end{figure}

We present  an objective diagnostic for the early identification of instabilities based on the maxima in finite-time changes of the curvature of material lines.
The maxima  are  precursors of significant wrinkling of material lines as
illustrated for a shear flow in figure \ref{fig:material_line}(a--d): a line of Lagrangian fluid tracers initialized along an unstable shear layer undergoes an oscillatory motion induced by an underlying velocity mode and rolls up along the forming vortices as the flow develops. Strong bending locations of material lines point to potential break-up locations  of the laminar flow field. These  can be visualized and identified through curvature changes. Figure \ref{fig:material_line} also illustrates that
the instability definition as a \textit{non-wall-bounded  wrinkling}  is distinct from the theory of the backbone of separation \citep{serra18}. 
While the latter is based on material curvature also, the separation theory
considers a \textit{wall-bounded} flow and, moreover, it identifies
a \textit{one-sided, upwelling}  instead of wrinkling  of the material line.

The kinematic instability identification approach does not rely on assumptions or knowledge of the averaged solution of the flow, as the diagnostic only depends on the kinematics of the material lines. 
The finite-time approach has practical benefits since
in experiments only limited data are available.
In addition to being objective, the finite-time integration of Lagrangian particles can  prove beneficial in the detection of intermittent instabilities, such as turbulent burst in a sub-critical Poiseuille flow, which will leave a footprint in the Lagrangian deformations even if the flow relaminarizes.
More generally, our approach can capture non-normal growths \citep{SH02}, an essential trigger for nonlinear behavior, which remain hidden to LSA.

We first provide analytic formulas connecting characteristic quantities used in LSA of perturbed shear flow with the induced flow map and Lagrangian curvature change. Then, we illustrate and verify the performance of the Lagrangian instability
identifier on three two-dimensional Navier-Stokes flows.
In a temporally developing jet flow, it is shown
that maximum curvature growth rates recover the linear growth rates of several unstable modes as predicted by LSA theory.
The unsteady, separated flow over a cambered airfoil at a moderate
Reynolds number of 20,000 shows that the curvature modes capture   instability
in the separated shear layer promptly over finite time without knowledge of a mean or base flow. 
Finally,  the approach is shown to 
identify instabilities in a flow without a mean
during the onset of wake instability in the growing wake behind a circular cylinder at low Reynolds number. 
In the following sections, the governing equations and the  numerical setup of the three cases are presented first (\S \ref{methodology}). Then,  the results are discussed in \S \ref{results}. Conclusions are reserved for \S \ref{conclusion}.

\section{Methodology and Setup} \label{methodology}
\subsection*{The curvature scalar}
\begin{figure}
    \centering 
	\includegraphics[width=0.8\textwidth]{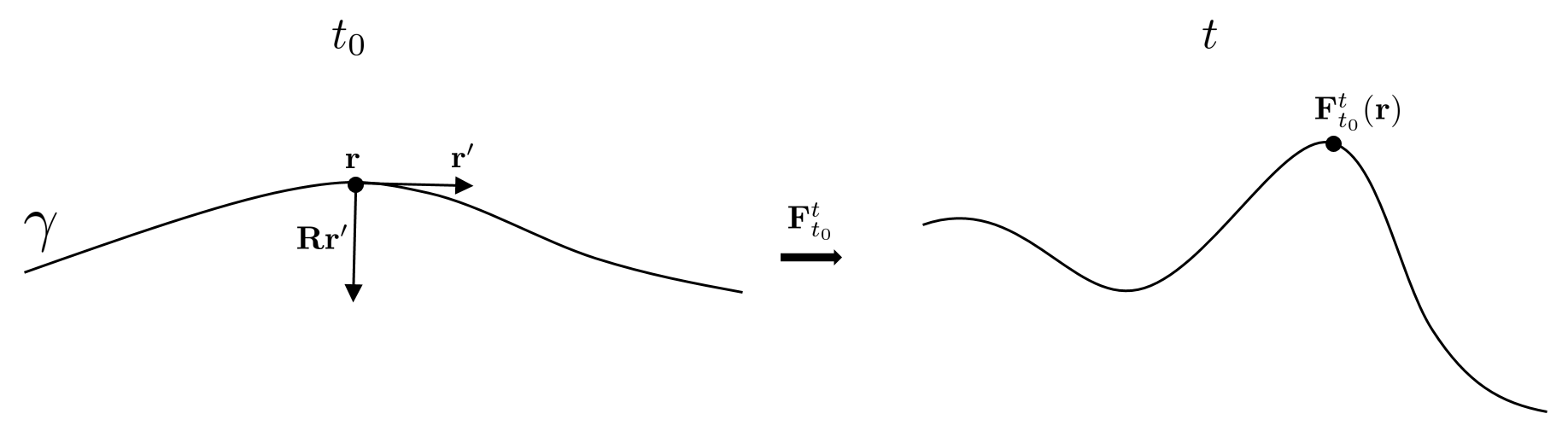}
	\caption{Sketch of a material line $\gamma$ parametrized at $t_0$ in the form $\mathbf{r}(s)$, $s\in[s_1,s_2]\subset \mathbb{R}$, transported and deformed by the flow map $\mathbf{F}_{t_0}^{t}$. At any point $\mathbf{r}(s)$, the Lagrangian folding  $\bar{\kappa}_{{t_0}}^{t_0+T}$ = $\kappa^{t_0+T}_{t_0}$ - $\kappa_0$ of $\gamma$ can be computed from equation \eqref{eq:kappa}.}
	\label{fig:FoldingIllustr}
\end{figure}

Following \citet{serra18}, we consider a smooth material curve $\gamma\in \mathbb{R}^2$, parametrized at $t_0$ in the form $\mathbf{r}(s)$, $s\in[s_1,s_2] \subset \mathbb{R}$, and denote its local tangent vector by $\mathbf{r^\prime}(s)$ and curvature scalar by $\kappa_0(s)=\tfrac{\langle \mathbf{r^{\prime \prime}}(s),\mathbf{Rr^\prime}(s)\rangle}{\sqrt{\langle \mathbf{r^{ \prime}}(s),\mathbf{r^{ \prime}}(s)\rangle}^3 }$ (see figure \ref{fig:FoldingIllustr}). The curvature change (or folding)  $\bar{\kappa}_{{t_0}}^{t_0+T}$ := $\kappa^{t_0+T}_{t_0}$ - $\kappa_0$ of $\gamma$ under the action of $\mathbf{F}_{t_{0}}^t$ can then be computed as  

\begin{multline} \label{eq:kappa}
\bar{\kappa}_{t_0}^{t_0+T} = \frac{\left\langle\left(\nabla^2\mathbf{F}_{t_0}^{t_0+T}(\mathbf{r})\mathbf{r}'\right)\mathbf{r}', \mathbf{R}\nabla\mathbf{F}_{t_0}^{t_0+T}(\mathbf{r})\mathbf{r}'\right\rangle}{\left\langle\mathbf{r}', \mathbf{C}_{t_0}^{t_0+T}(\mathbf{r})\mathbf{r}'\right\rangle^{3/2}} +
\kappa_{0}\left[\frac{\det\left(\nabla\mathbf{F}_{t_0}^{t_0+T}(\mathbf{r})\right)\left\langle\mathbf{r}',\mathbf{r}'\right\rangle^{3/2}}{\left\langle\mathbf{r}', \mathbf{C}_{t_0}^{t_0+T}(\mathbf{r})\mathbf{r}'\right\rangle^{3/2}} - 1\right],
\end{multline}
where $\langle \cdot , \cdot \rangle$ denotes the inner product; $(\nabla^2 \mathbf{F}_{t_0}^{t}(\mathbf{r})\mathbf{r}')_{ij}=\sum\limits_{k=1}^{2}{\partial_ {jk}F_{t_{0}}^t}_{i}(\mathbf{r})r'_k,\ i,j\in\{1,2\}$, $\mathbf{R}$ is a clockwise ninety-degree rotation matrix defined as
\begin{align}\label{eq:R}
\mathbf{R} := \begin{bmatrix}
0 & 1 \\
-1 & 0
\end{bmatrix},
\end{align}
and $\mathbf{C}_{t_0}^{t}$ = $[\nabla\mathbf{F}_{t_0}^{t}]^{\top}\nabla\mathbf{F}_{t_0}^{t}$ is the right Cauchy-Green strain tensor. Equation \ref{eq:kappa} shows that the Lagrangian folding depends on spatial inhomogeneities of $\mathbf{F}_{t_0}^{t}$, the curve stretching encoded by $\mathbf{C}_{t_0}^{t}$, the initial curvature and the compressibility of the flow described by $\det\left(\nabla\mathbf{F}_{t_0}^{t}\right)$.
See \citet{serra18} for a more detailed description. 

\subsection*{Linear stability analysis}
Instabilities in shear flows are commonly analyzed by the growth of a perturbation to a  base flow.
The incompressible flow field described by $u$, $v$, and $p$ is decomposed into a base with a perturbation as $u$ = $U$ + $u'$, $v$ = $V$ + $v'$, and $p$ = $P$ + $p'$, which is then substituted into the Navier-Stokes equations. 
Assuming small, exponentially growing perturbations of the form
\begin{equation}
\label{eq:pert}
\left[u',v',p'\right] = \operatorname{Re}\left([\hat{u}(y),\hat{v}(y),\hat{p}(y)] e^{i(kx - \omega t)}\right),
\end{equation}
the governing equations can be linearized and solved as an eigenvalue problem (see \citet{drazin_2003} for details). In \eqref{eq:pert}, the quantities $\hat{u}(y)$, $\hat{v}(y)$, and $\hat{p}(y)$ are the complex amplitudes of the perturbations and $k$ and $\omega$ are the real wavenumber and complex frequency. 
For the analysis of a temporal developing flow, we consider temporal growth governed by the imaginary part $\omega_i$ of the complex frequency.
An eigenvalue solver determines the frequencies $\omega_i$ and corresponding eigenmodes $\hat{u}$, $\hat{v}$, and $\hat{p}$ for a given wavenumber $k$.

\subsection*{Lagrangian Curvature growth}
A  theoretical connection between linear stability analysis 
and the curvature analysis can be made for material lines initialized along the \textit{x}-axis such that $\mathbf{r}'$ = $[1, \, 0]^{\top}$. For these lines, the material time derivative of $\bar{\kappa}_{t_{0}}^{t_{0}+T}$ relates to the lateral velocity as $\dot{\kappa}_{t_{0}}$ = $-\partial_{xx}v$ 
as proven in appendix \ref{appendix_kappa}. 
Decomposition of $v$ into a base flow $V$ and fluctuating component $v'$ yields $\dot{\kappa}_{t_0}$ = $-(\partial_{xx}v' + \partial_{xx}V)$.
With the base flow $V$ = 0 for the planar jet flow, the \textit{y}-velocity component is $v$ = $v'$ = $\hat{v}(y)e^{ikx - i\omega t}$ and substitution results in
\begin{equation}
\label{eq:kdot}
\dot{\kappa}_{t} = -\partial_{xx}v' = -\hat{v}(y)\partial_{xx}e^{ikx - i\omega t} = k^2\hat{v}(y)e^{ikx - i\omega t} = k^2 v',
\end{equation}
showing that the rate-of-change of $\kappa$ scales with $v'$ by the square of the wavenumber. This connection, however, is valid only instantaneously and hence lacks the memory trace, or Lagrangian history, of fluid flows.  

To understand how perturbations of a base flow in the infinite-dimensional space leave a footprint on the physical phase space of fluid particles, we derive an analytic expression relating the Eulerian growth rate and wavenumber of the perturbation and the induced Lagrangian folding of fluid. We focus our analysis on general parallel shear flows perturbed with small perturbations of the form $\mathbf{u_{\epsilon}} = \mathbf{u_0}+\epsilon\mathbf{u^\prime}$, $0<\epsilon\ll 1$, where $\mathbf{u_0}=[U(y), 0]^\top$ is the base flow and $\mathbf{u^\prime}=\operatorname{Re}([\hat{u}(y),\hat{v}(y)]e^{i(kx - \omega t)})$ the perturbation . Because there are no analytic flow maps $\mathbf{F_{\epsilon}}$ induced by $\mathbf{u_{\epsilon}}$, we first find an analytical approximation of $\mathbf{F_{\epsilon}}$ neglecting $\mathcal{O}(\epsilon^2)$ terms. We then use the approximated $\mathbf{F_{\epsilon}}$ and equation \eqref{eq:kappa} to compute an analytical expression of the Lagrangian curvature change for initially straight material lines. Using the formulas derived in appendix {\ref{appendix_kbar}}, we obtain the following result.

\begin{thm}\label{sec:Thm1}
	Consider a planar parallel shear flow of the form $\mathbf{u_{\epsilon}} = \mathbf{u_0}+\epsilon\mathbf{u^\prime}$, $0<\epsilon\ll 1$, where $\mathbf{u_0}=[U(y), 0]^\top$ is the base flow and $\epsilon \mathbf{u^\prime}=\epsilon \operatorname{Re}([\hat{u}(y),\hat{v}(y)]^{\top}e^{i(kx - \omega t)})$ the perturbation. The flow map $\mathbf{{{_\epsilon}F}}_{t_0}^{t_0+T}(\mathbf{x_0})$ induced by $\mathbf{u_{\epsilon}}$ admits an analytic solution at leading order, of the form
	\begin{equation}\label{eq:F_Thm}
    \mathbf{{{_\epsilon}F}}_{t_0}^{t_0+T}(\mathbf{x_0})=\mathbf{{{_0}F}}_{t_0}^{t_0+T}(\mathbf{x_0}) + \mathbf{A}_{t_{0}}^{t_0+T}(\mathbf{x_0})\epsilon + \mathcal{O}(\epsilon^2),
    \end{equation}
    where $\mathbf{A}_{t_{0}}^{t_0+T}(\mathbf{x_0})$ satisfies the following initial value problem \begin{equation}
	\begin{cases}
		\dot{\overline{\mathbf{A}_{t_{0}}^{t}}}(\mathbf{x_0})=\mathbf{\nabla u_0}(\mathbf{{{_0}F}}_{t_0}^t(\mathbf{x_0}))\mathbf{A}_{t_{0}}^{t}(\mathbf{x_0}) + \mathbf{{u^\prime}}(\mathbf{{{_0}F}}_{t_0}^{t}(\mathbf{x_0}),t)\\
		\mathbf{A}_{t_{0}}^{t_0}(\mathbf{x_0}) = \mathbf{0}\\
		\mathbf{{{_0}F}}_{t_0}^{t}(\mathbf{x_0}) = \mathbf{x_0} + \mathbf{u_0}(\mathbf{x_0})(t-t_0).
	\end{cases}
	\label{eq:ADiffeq}
\end{equation}
We provide the explicit formula for $\mathbf{{{_\epsilon}F}}_{t_0}^{t_0+T}(\mathbf{x_0})$ in equation \eqref{eq:flowmap_eps}.

	The curvature change induced by $\mathbf{u_{\epsilon}}$ during $[t_0,t_0+T]$ on an initially straight ($\kappa_0=0$) material line located at $\mathbf{x_0}=[x_0,y_0]^{\top}$ and aligned with the stream-wise direction $\mathbf{r^{\prime}}=[1,0]^{\top}$, can be computed as
\begin{equation}\label{eq:kappa_barThm}
\begin{split}
    \bar{\kappa}_{t_0}^{t_0+T} (\mathbf{x_0})&=
    \frac{k^2}{\omega_i^2 + \omega_r^2 - 2 \omega_r k U(y_0) + k^2 U^2(y_0)}\bigg[\\
        &e^{\omega_i t_0}\Big(
        \big(-\omega_i\hat{v}_r(y_0)+(\omega_r-kU(y_0))\hat{v}_i(y_0)\big)\cos(kx_0-\omega_r t_0) \\
        &+\big(\omega_i\hat{v}_i(y_0)+(\omega_r-kU(y_0))\hat{v}_r(y_0)\big)\sin(kx_0-\omega_r t_0)
        \Big) \\
        &+e^{\omega_i(t_0+T)}\Big(
        \big(\omega_i\hat{v}_r(y_0)-(\omega_r-kU(y_0))\hat{v}_i(y_0)\big)\cos(kx_0-\omega_r(t_0+T)+kU(y_0)T) \\
        &-\big(\omega_i\hat{v}_i(y_0)+(\omega_r-kU(y_0))\hat{v}_r(y_0)\big)\sin(kx_0-\omega_r(t_0+T)+kU(y_0)T)
        \Big)
    \bigg] + \mathcal{O}(\epsilon^2),
\end{split}
\end{equation}
where $\hat{u}=\hat{u}_r + i \hat{u}_i,\ \hat{v}=\hat{v}_r + i \hat{v}_i$ are complex amplitudes,  $\omega=\omega_r + i\omega_i$ is the complex frequency and $k$ the real wave number. \\
\noindent Proof. $\mathrm{See\ appendix\ \ref{appendix_kbar}.}$ \hfill\ensuremath{\square}
\end{thm}

\subsection*{Properties of Lagrangian Curvature growth}
\label{properties}
Equation \eqref{eq:kappa_barThm} provides an analytic relation between a velocity field composed of normal modes and the Lagrangian curvature change and therefore directly connects LSA-based quantities with fluid material wrinkling, which is objective under \eqref{eq:obj}.
Several essential properties of the curvature change in unstable flows given by equation \eqref{eq:kappa_barThm} follow:
\vspace{1em}
\begin{enumerate}[(i)]
 \setlength\itemsep{1em}
\item First and foremost, \textit{the curvature change grows exponentially at a rate $\omega_i$ according to LSA} after adjusting from the initial, zero $\bar{\kappa}$. This  theoretical proof solidifies the validity of the definition of the instability through curvature modes and connects it with 
the LSA for  velocity fields which is not frame-invariant.
\item For $\omega_i$ < 0 (ceasing perturbation), the $\bar{\kappa}$ converges to a finite value because the initial perturbation is retained in the fluid deformation (cf. figure \ref{fig:kt0TLOT}). This memory trace of fluid history is particularly relevant in the context of non-normal growth.
\item The curvature and the velocity field are phase shifted if the fluid particles move relative to the velocity perturbation mode.
Along the \textit{critical layer} \citep{drazin_2003}, i.e. where $U$ = $c_r$ = $\omega_r/k$, the curvature maxima of material lines travel in phase with the perturbation and are therefore material.
\end{enumerate}


\section{Test cases}
Solutions to  the two-dimensional compressible Navier-Stokes equations for conservation of mass, momentum and energy in non-dimensional form are obtained
using  a discontinuous Galerkin spectral element method
for spatial approximation and a 4\textsuperscript{th}-order explicit Runge-Kutta scheme for temporal integration. For a description of the method
and validation of the solver we refer the interested reader
to \citet{Kopriva09,nelson15} and references therein. 
Lagrangian fluid particles are traced with a 3\textsuperscript{rd} order Adam-Bashfort scheme in time according to \citet{MDF19}.

\subsection*{Temporarily developing jet}
\begin{figure}
    \makebox[\textwidth][c]{
	\includegraphics[width=0.2\textwidth]{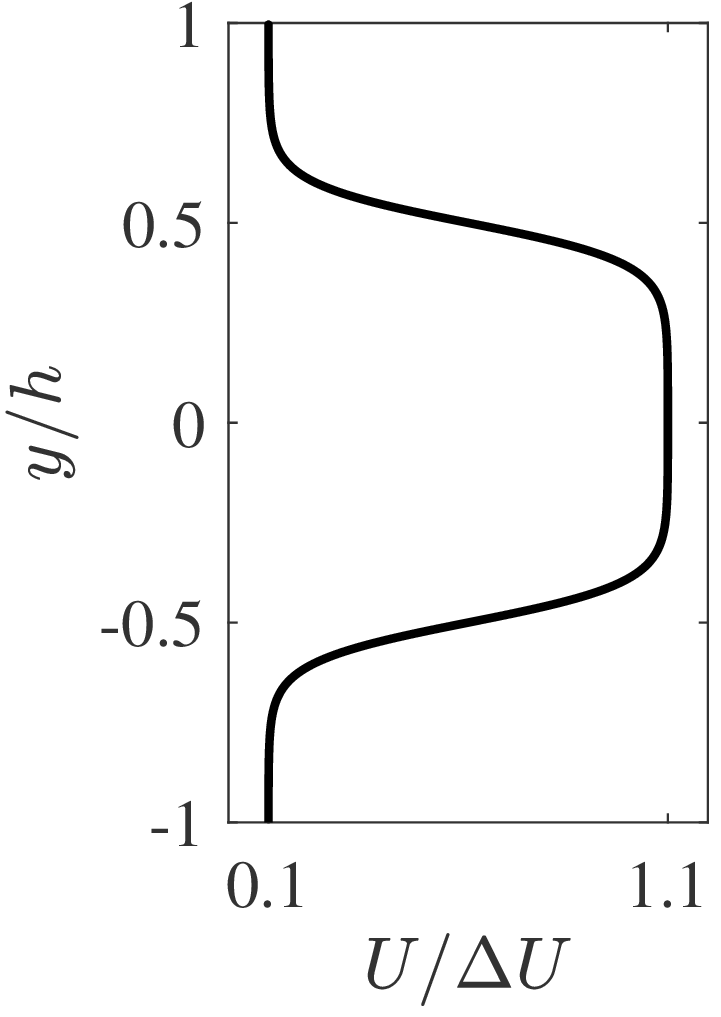}
	\hfill
	\includegraphics[width=0.3\textwidth]{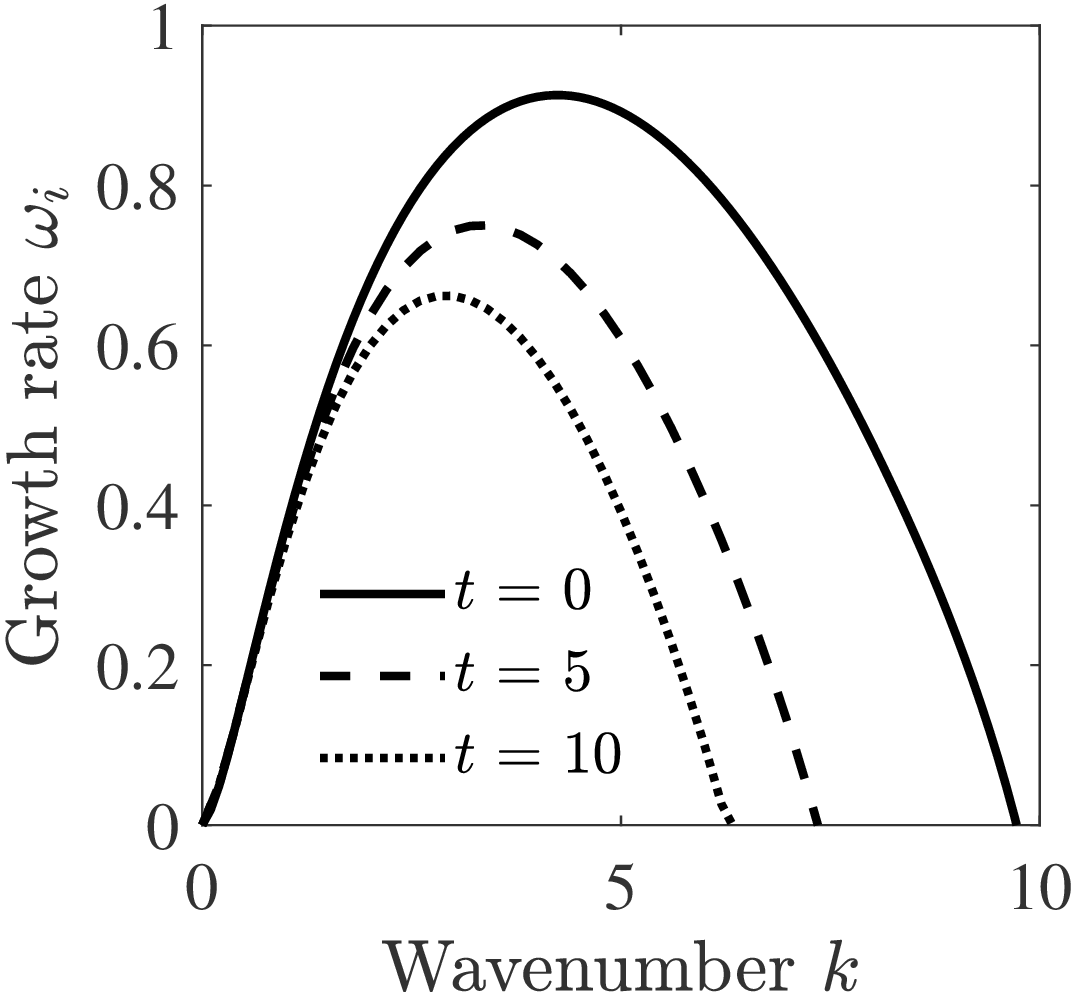}
	\hfill
	\includegraphics[width=0.35\textwidth]{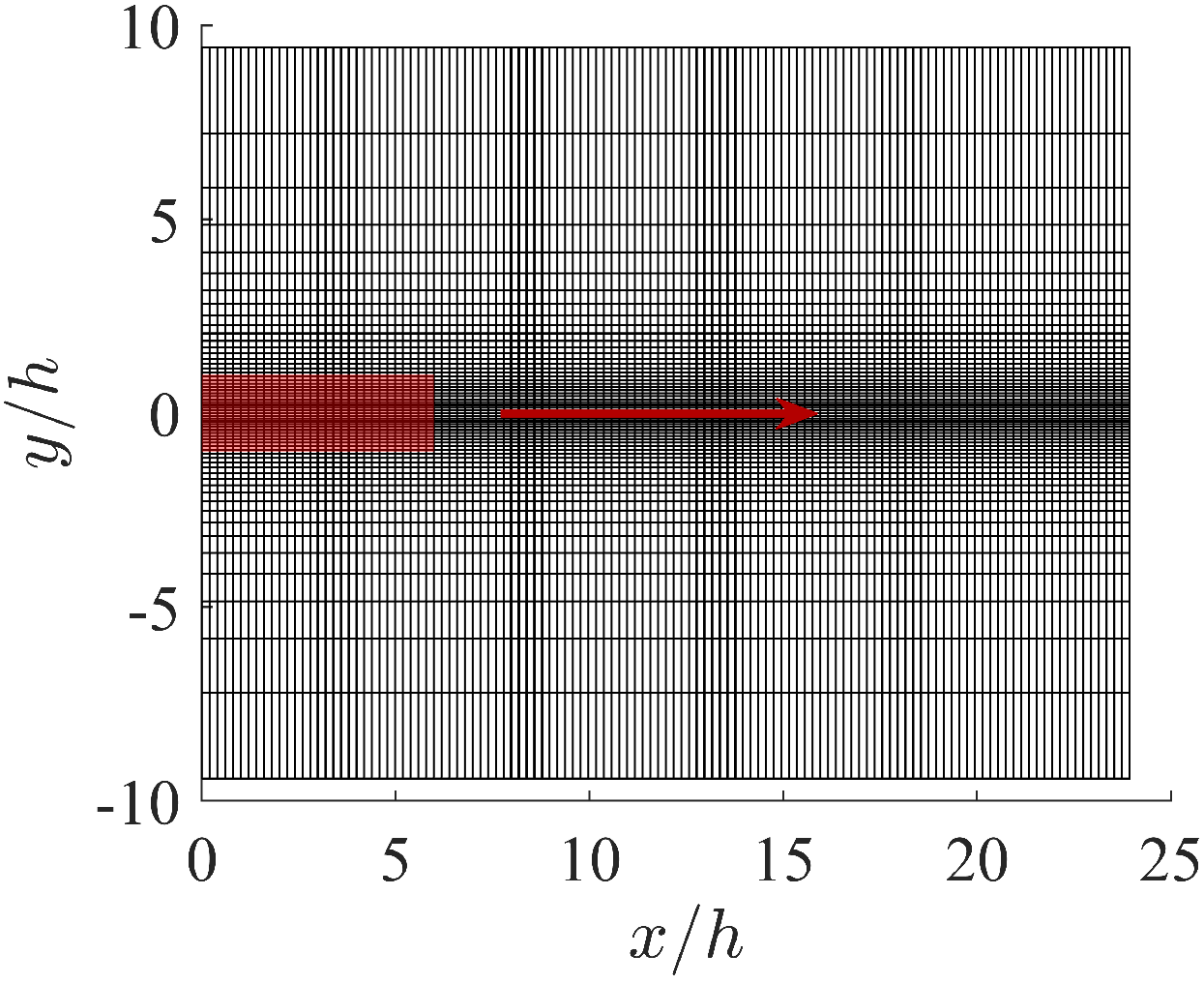}
	}
	\makebox[\textwidth][c]{
    \makebox[0.2\textwidth][c]{(a) Velocity profile}
    \hfill
    \makebox[0.3\textwidth][c]{(b) Dispersion relation}
    \hfill
    \makebox[0.35\textwidth][c]{(c) Computational domain}
    }
	\caption{(a) Hyperbolic tangent, jet flow  velocity profile normalized by the velocity difference $\Delta U$ between center and coflow. (b) Dispersion relation, i.e. temporal growth rates $\omega_i(t)$ versus wavenumbers $k$, determined with LSA
	for viscous jet velocity profiles at $t$ = 0, 5, and 10. (c) Computational domain and grid in normalized coordinates with initial particle positions (shaded red area) and flow direction (red arrow). }
	\label{fig:baseflow}
\end{figure}

The temporal stability of a top-hat jet profile is investigated computationally in a periodic domain. The characteristic scales of the problem are the jet width $h$ and the velocity difference $\Delta U$ between center flow $U_1$ and coflow $U_2$. The field is initialized with a  hyperbolic tangent function velocity profile  (see figure \ref{fig:baseflow}a),
\begin{equation}
\label{eq:init}
U = \frac{U_1-U_2}{2}\left( \tanh\frac{y+\frac{1}{2}h}{2\theta} - \tanh\frac{y-\frac{1}{2}h}{2\theta}\right) + U_2,
\end{equation}
where the lateral velocity component is $V$ = 0. The center and coflow velocities are $U_1/\Delta U$ = 1.1 and $U_2/\Delta U$ = 0.1, respectively. 

Following \citet{stanley02},  the shear layer momentum thickness is set to $\theta/h$ = 1/20 and the Reynolds number based on the jet width $h$ and velocity difference $\Delta U$ to $Re_h$ = $\Delta U h/\nu$ = 3000. A free-stream Mach number of $M$ = 0.1, based on a characteristic velocity of unity ($M$ = 0.11 if based on the center flow and $M$ = 0.01 if based on the outer flow),
ensures that flow has negligible compressibility effects. This is verified
by a comparison of a  simulation at  $M$ = 0.05 which shows no discernible changes in the results presented below.

\begin{figure}
	\makebox[\textwidth][c]{
	\includegraphics[width=0.3\textwidth]{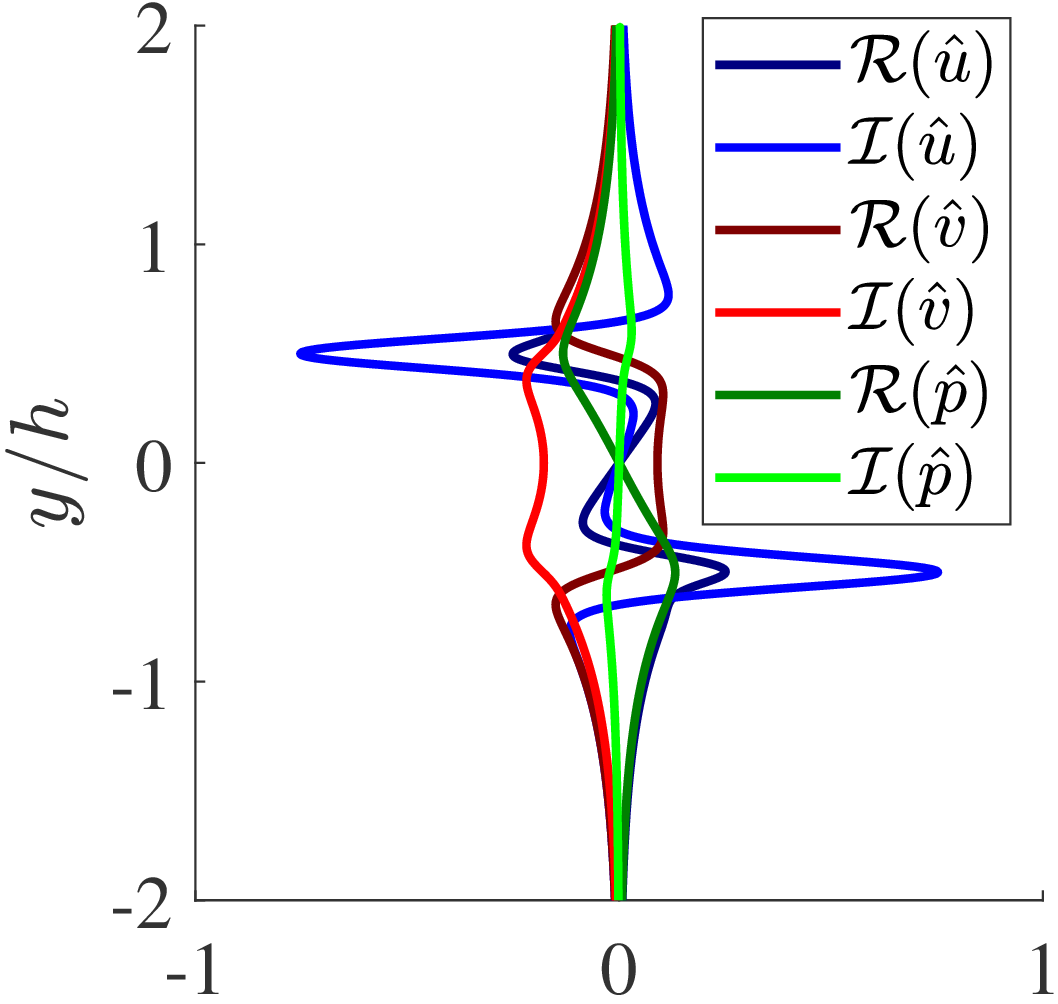}
	\hfill
	\includegraphics[width=0.3\textwidth]{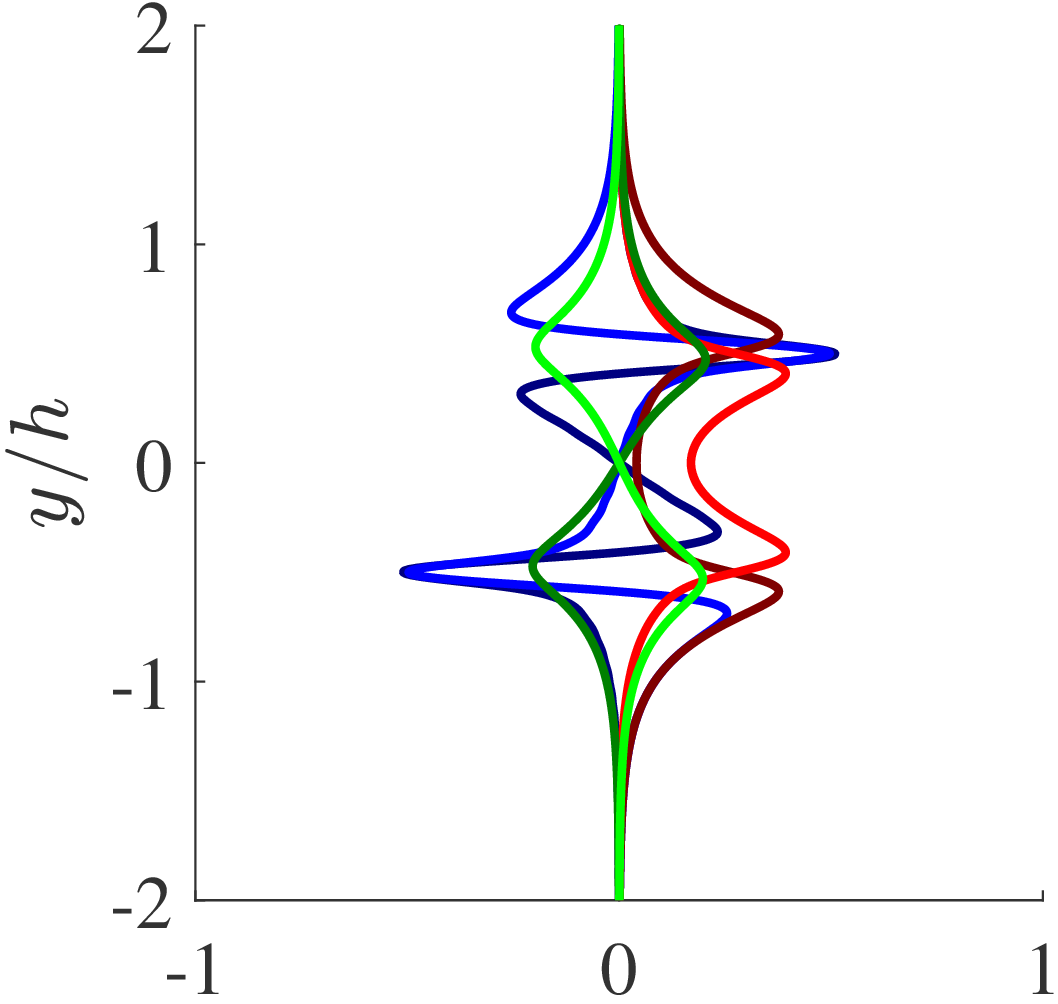}
	\hfill
	\includegraphics[width=0.3\textwidth]{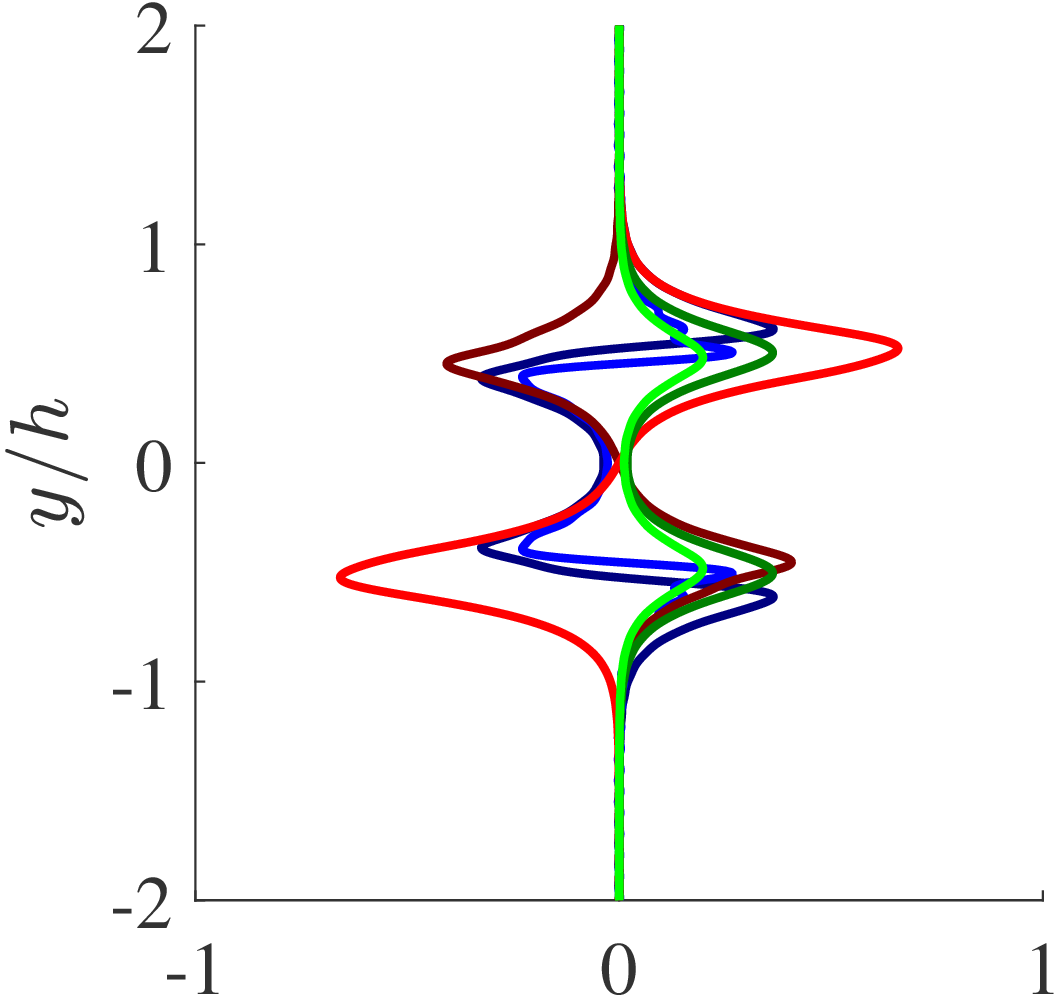}
	}
	\makebox[\textwidth][c]{
	\makebox[0.3\textwidth][c]{(a) $k$ = 2.1}
	\hfill
	\makebox[0.3\textwidth][c]{(b) $k$ = 4.2}
	\hfill
	\makebox[0.3\textwidth][c]{(c) $k$ = 8.4}
	}
	\caption{Real and imaginary components of the unstable eigenmodes of $u$, $v$, and $p$ obtained from LSA for wavenumbers $k$ = 2.1 (a), $k$ = 4.2 (b), and $k$ = 8.4 (c).}
	\label{fig:modes}
\end{figure}

Using LSA, the  dispersion relation is determined for the incompressible jet flow profile. It is plotted in figure \ref{fig:baseflow}(b) and peaks at a wavenumber of $k_{max}$ = 4.2. The eigenmodes corresponding to wavenumbers $k$ = 2.1, $k$ = 4.2 and $k$ = 8.4 are visualized in figure \ref{fig:modes}(a--c). 
Note that the temporarily developing viscous jet flow does not have a time-independent base flow state and diffuses in time. 
The dipersion relation as shown in figure \ref{fig:baseflow}(b) is therefore time-dependent also.
Based on the wavelength of this most unstable mode $\lambda_{max}$ = $2\pi/k_{max}$ at $t$=0, the computational domain is sized to fit 16 waves in \textit{x}-direction ($x$ $\in$ $[0, \, 16\lambda_{max}]$) and is large enough in \textit{y}-direction ($y/h$ $\in$ $[-3\pi, \, 3\pi]$) for the eigenmodes (see figure \ref{fig:modes}) to reduce to machine precision zero.
To avoid grid resolution issues, an unnecessarily high number of  7200 quadrilateral elements clustered around the center line is used with the solution approximated with a 9\textsuperscript{th} order polynomial within each element (see figure \ref{fig:baseflow}c). Periodic boundary conditions are specified in $x$-direction and Riemannian free-stream boundary conditions \citep{JKM03} are imposed in $y$-direction. The velocity field is initialized according to \eqref{eq:init}.
The real part of the eigenmodes corresponding to $k_{pert}$ = 2.1, $k_{pert}$ = $k_{max}$ = 4.2, and $k_{pert}$ = 8.4 (see figure \ref{fig:modes}) are scaled with a factor of $10^{-3}$ and superimposed onto the base flow as initial perturbation. Additionally, we also consider random noise perturbations.
A sheet of 501$\times$251 Lagrangian fluid tracers is initialized in the region $x_p$ $\in$ $[0, \, 4\lambda_{max}]$ and $y_p$ $\in$ $[-h, \, h]$ for cases using LSA eigenmodes and 2001$\times$251 ($x_p$ $\in$ $[0, \, 16\lambda_{max}]$) tracers are initialized for computations with random noise perturbations.

\subsection*{NACA 65(1)-412 airfoil flow}

To test the kinematic approach for an unstable
flow without an analytical base field, we
consider the stability of a separated and unstable shear layer 
in the flow field over a cambered NACA 65(1)-412 airfoil at 7$^\circ$ incidence 
for a chord-based Reynolds number of 
$Re_c$ = 20,000 and a free-stream Mach number of $M$ = 0.3. 
The parameter $c$ refers to the chord length of the wing. 
We have studied flow separation and Lagrangian wake dynamics
extensively for this geometry in previous work 
(e.g. \cite{nelson16,KJTS18,TKJS19a}) and use
the setup that yields a grid-independent solution 
with negligible blockage effects of these investigations here.

The computational domain (C-grid) consists of 23,400 quadrilateral elements and, within each element, the solution is approximated by 6\textsuperscript{th} order Legendre-Gauss polynomials. 
The in- and outflow boundaries of the domain are 30 chord lengths away from the airfoil, 
yielding a  blockage effect of 1\% by the computational domain. 
A damping layer at the outflow boundary reduces pressure reflections from outflow boundary \citep{JKM03}.
The mesh is refined around the airfoil (see figure \ref{fig:airf_cyl_mesh}a), with  first grid point away from the wall at approximately $y_{g}^+$ = 0.2. 
where, $y^+$ refers to the dimensionless wall coordinate.
Figure \ref{fig:airf_cyl_mesh}(a) further illustrates the location of the 4001$\times$300 Lagrangian fluid tracers that are initialized in wall-parallel lines along the upper surface of the airfoil.

\subsection*{Circular cylinder}

In a final test, the origin and onset of vortex shedding in the wake  behind circular cylinder at $Re_d$ = 200 is considered 
with the goal of illustrating
the objective nature of the kinematic approach.
\citet{SS90} show  that the vortex shedding behind a circular cylinder
has its origin in unstable modes in the growing cylinder wake region.
Control of those modes 
can delay or prevent shedding from occurring.
Because the instability is associated with temporal growth of the wake, a steady base velocity and a corresponding
velocity perturbation cannot be defined. The finite-time material line
curvature change  can be.

\begin{figure}
    \makebox[\textwidth][c]{
	\includegraphics[width=0.5\textwidth]{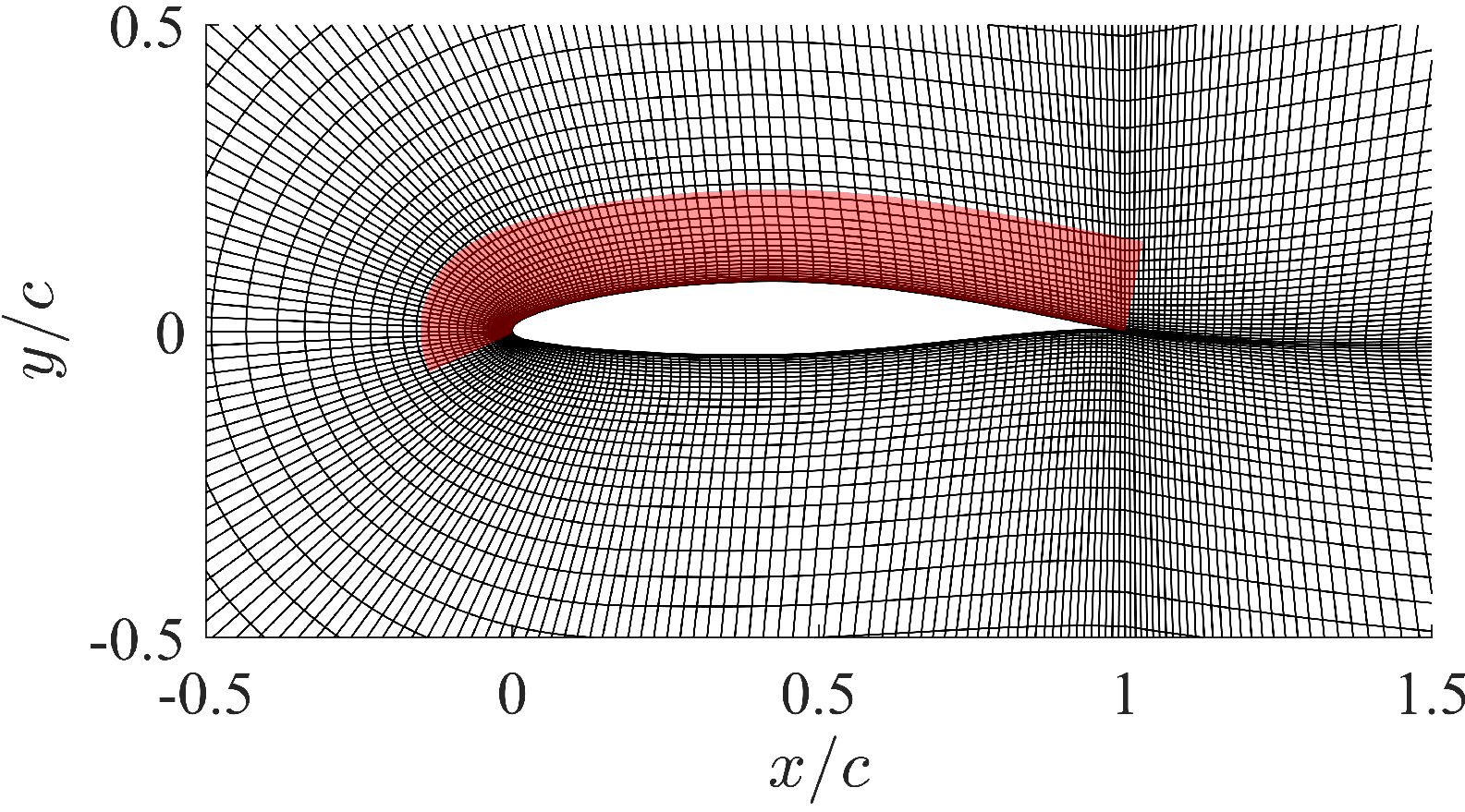}
	\hfill
	\includegraphics[width=0.36\textwidth]{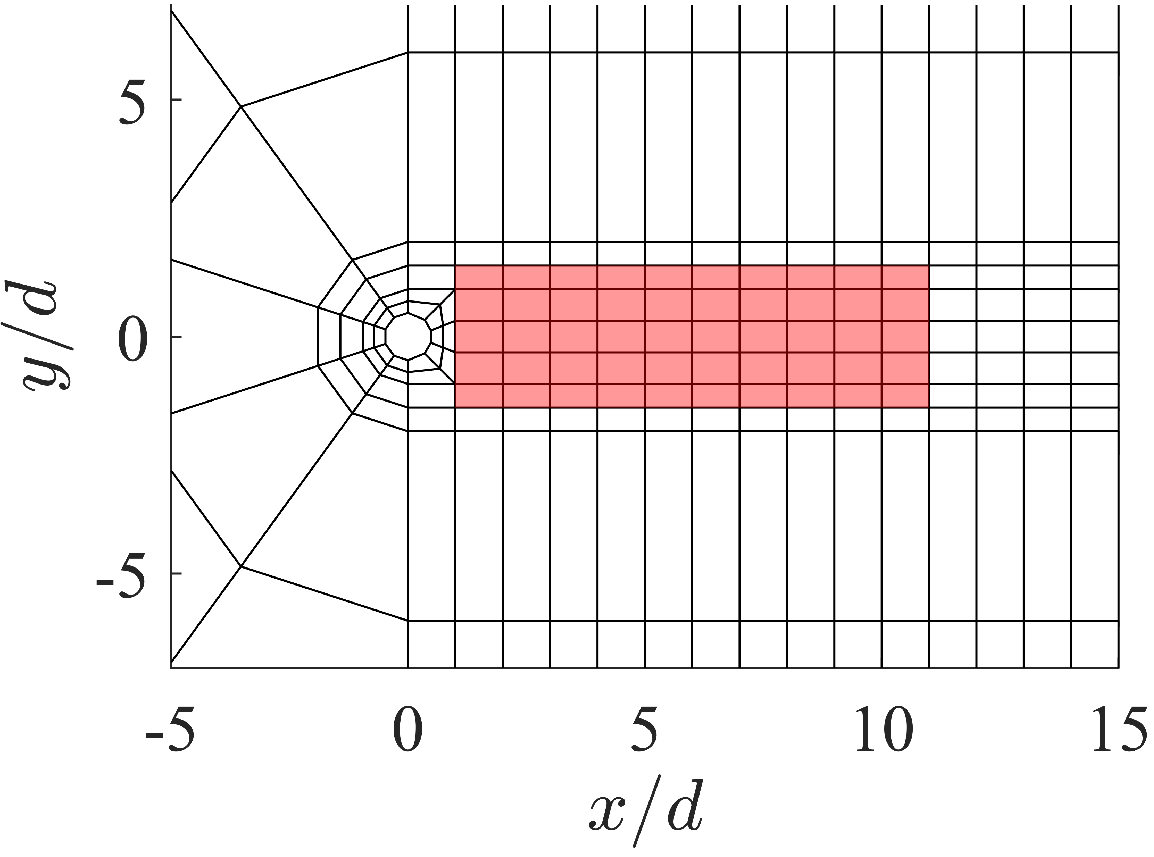}
	}
	\makebox[\textwidth][c]{
	\makebox[0.5\textwidth][c]{(a) NACA 65(1)-412}
	\hfill
	\makebox[0.36\textwidth][c]{(b) Circular cylinder}
	}
	\caption{Computational domain and grid used for simulations of the flows over (a) a NACA 65(1)-412 airfoil and (b) a circular cylinder.  	Only elements edges (no quadrature grid points within the element)   are shown.
	Fluid particle tracers are initialized in the element colored in red.}
	\label{fig:airf_cyl_mesh}
\end{figure}

Following the problem setup in \citet{KSJ20a} that was used
for the analysis of  kinematic aspects of Lagrangian separation,
a free-stream Mach number of $M$ = 0.1 is specified.
The grid-independent solution \citep{KSJ20a} is approximated with a high, 18\textsuperscript{th} order  polynomial approximation on 347 quadrilateral elements (see figure \ref{fig:airf_cyl_mesh} b).
Riemannian free-stream conditions are prescribed at all outer boundaries and the vertical domain size $y/d$ $\in$ $[-20, \, 20]$ yields a blockage of 2.5\%.
A sheet of 501$\times$251 Lagrangian fluid tracers is initialized half a diameter downstream from the cylinder in the region $x_p$ $\in$ $[1d, \, 11d]$ and $y_p$ $\in$ $[-1.5d, \, 1.5d]$, as shown in figure \ref{fig:airf_cyl_mesh}(b).

\section{Results and Discussion} \label{results}
\subsection*{Temporarily developing jet}
\begin{figure}
    \makebox[\textwidth][c]{
	\includegraphics[width=0.28\textwidth]{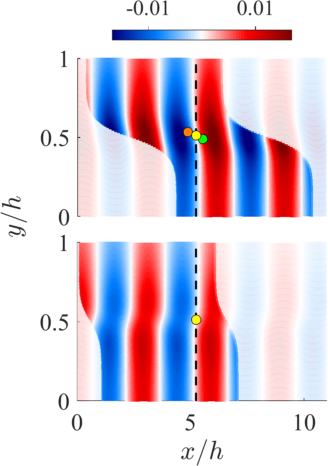}
	\hfill
	\includegraphics[width=0.305\textwidth]{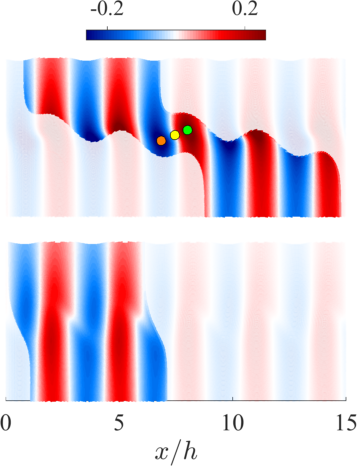}
	\hfill
	\includegraphics[width=0.378\textwidth]{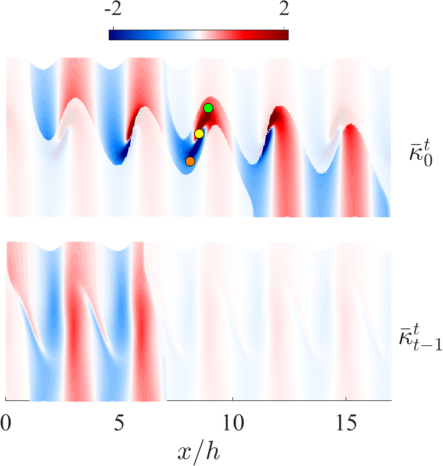}
	}
	\makebox[\textwidth][c]{
    \makebox[0.28\textwidth][c]{(a) $t$ = 4.0}
    \hfill
    \makebox[0.305\textwidth][c]{(b) $t$ = 8.0}
    \hfill
    \makebox[0.378\textwidth][c]{(c) $t$ = 10.0}
	}
	\caption{Contours of the curvature scalar field $\bar{\kappa}_{0}^{t}$ (upper figures) and $\bar{\kappa}_{t-1}^{t}$ (lower figures) at $t$ = 4.0 (a), $t$ = 8.0 (b) and $t$ = 10.0 (c) for a jet flow computation with initial velocity perturbation according to eigenmodes with a wavenumber $k_{pert}$ = 2.1. 
	Only the upper half of the jet is shown.
	Time units are scaled by $h/\Delta U$.
	Bright visualizations are for directly computed fields. The faded contours are copies of the bright contours.
	Color-coded tracers: $\max(\vert \mathbf{\nabla}\bar{\kappa}_0^8\vert)$ in yellow, $\max(\bar{\kappa}_0^8)$ in green, and $\min(\bar{\kappa}_0^8)$ in orange, based on the scalar field in (b).
	Material position of color-coded tracers at earlier and later times indicated in (a) and (c).
	Dashed line indicates indicates location of center marker in upper figure and is duplicated in lower figure, together with yellow marker. }
	\label{fig:alpha_2p1}
\end{figure}

\begin{figure}
    \makebox[\textwidth][c]{
	\includegraphics[width=0.28\textwidth]{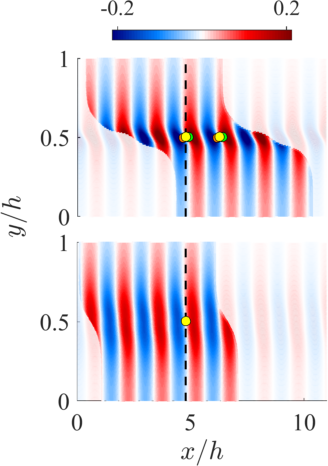}
	\hfill
	\includegraphics[width=0.258\textwidth]{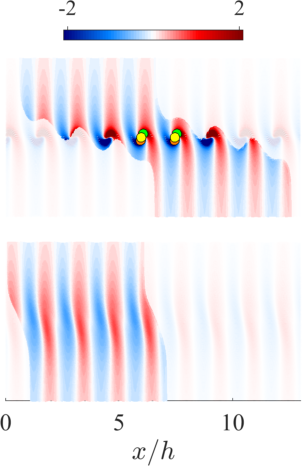}
	\hfill
	\includegraphics[width=0.343\textwidth]{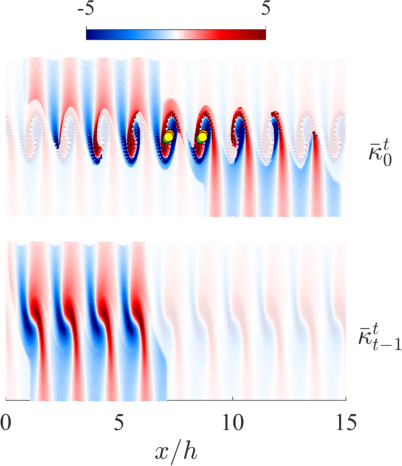}
	}
	\makebox[\textwidth][c]{
    \makebox[0.28\textwidth][c]{(a) $t$ = 4.0}
    \hfill
    \makebox[0.258\textwidth][c]{(b) $t$ = 6.0}
    \hfill
    \makebox[0.343\textwidth][c]{(c) $t$ = 8.0}
	}
	\caption{Contours of the curvature scalar field $\bar{\kappa}_{0}^{t}$ (upper figures) and $\bar{\kappa}_{t-1}^{t}$ (lower figures) for $t$ = 4.0 (a), $t$ = 6.0 (b) and $t$ = 8.0 (c) for a jet flow computation with initial velocity perturbation according to eigenmodes with a wavenumber $k_{pert}$ = 4.2. 
	Only the upper half of the jet is shown.
	Time units are scaled by $h/\Delta U$.
	Bright visualizations are for directly computed fields. The faded contours are copies of the bright contours.
	Color-coded tracers: $\max(\vert\mathbf{\nabla}\bar{\kappa}_0^6\vert)$ in yellow, $\max(\bar{\kappa}_0^6)$ in green, and $\min(\bar{\kappa}_0^6)$ in orange, based on the scalar field in (b).
	Material position of color-coded tracers at earlier and later times indicated in (a) and (c).
	Dashed line indicates indicates location of center marker in upper figure and is duplicated in lower figure, together with yellow marker. }
	\label{fig:alpha_4p2}
\end{figure}

Finite-time instabilities are extracted from Lagrangian particles by integrating fluid tracers from some time $t_0$ over the interval $T$ until the time $t$ = $t_0$ + $T$.
Given that the curvature change depends on both, $t_0$ and $T$, two cases are considered for the computation of $\bar{\kappa}_{t_0}^{t_0+T}$ based on \eqref{eq:kappa}: (a) the reference state is set at $t_0 = 0$ and we calculate $\bar{\kappa}_{0}^{t}$ based on a single particle trace or (b) the integration interval $T$ is kept constant and the curvature follows as $\bar{\kappa}_{t-T}^{t}$ with the particle trace re-initialized every interval $T$.
In both cases, the tangent vector is $\mathbf{r}'$ = $[1, 0]^T$ and $\kappa_0$ = 0, such that only the first term of \eqref{eq:kappa} contributes to the curvature change. 
The resulting curvature scalar field $\bar{\kappa}_{t_0}^{t_0+T}$ is plotted over the advected particle positions in figures \ref{fig:alpha_2p1} and \ref{fig:alpha_4p2}. 

Consistent with figure \ref{fig:material_line}, the time series (a) -- (c) shows how the initial waviness in the curvature field and 
associated early wrinkles in the material line are precursors
of regions with  increased and concentrated curvature change  magnitudes
associated with vortex roll-up at later times.
Local extrema in the curvature fields (green and orange markers in figures \ref{fig:alpha_2p1} and \ref{fig:alpha_4p2}), as well as the point 
in between with a maximum gradient (yellow marker), represent  point identifiers of significant material wrinkling and vortex formation at later times.
Per the property (iii) described in section \ref{properties}, local maxima of the Lagrangian curvature change in unstable, parallel shear layers are only material along the critical layer where $U(y)$ = $c$ and can shift elsewhere in regions where the unstable mode has a different phase
velocity as compared to the flow's convection velocity.
The yellow point markers are located exactly on the critical lines at
$y$ = $\pm$0.51$h$ and $y$ = $\pm$0.50$h$  for $k_{pert}$ = 2.1 and $k_{pert}$ = 4.2
and can be considered critical points in the Lagrangian analysis.
\makeatletter
\let\@float@original\@float
\xpatchcmd{\@float}{\csname fps@#1\endcsname}{h!}{}{}
\makeatletter
\let\@float\@float@original
\makeatother

The shift between the vertical curvature change bands  visualized in Figure \ref{fig:alpha_2p1} and \ref{fig:alpha_4p2}  for  long and short integration intervals, i.e. $\bar{\kappa}_{0}^{t}$ and $\bar{\kappa}_{t-1}^{t}$, respectively, is also explained  by the relative velocity difference
of the modes and the fluid.
Because the phase shift between the finite-time curvature change and the velocity mode depends on the integration interval $T$ of the particle trace (cf. equation \eqref{eq:kappa_barThm}), different initial times $(t-T)$ of $\bar{\kappa}_{t-T}^t$ yield a different phase in the curvature field.
Along the critical layer, particles are not subjected to passing unstable wave modes,
and the short and long time-integrated curvature contours have their maxima and minima at the same location. 
Above and below, the fluid particles move at a different speed as compared to the velocity mode such that the bands are slightly shifted with respect to each other.
An analysis of the phase shift for a simple traveling mode is presented  with additional visualizations in appendix \ref{appendix_wave}
to provide  a theoretical basis for the phase shift.

\begin{figure}
	\makebox[\textwidth][c]{
	\makebox[0.03\textwidth][l]{\rotatebox{90}{\hspace{20pt} $k_{pert}$ = $2.1$}}
	\hfill
    \includegraphics[width=0.25\textwidth]{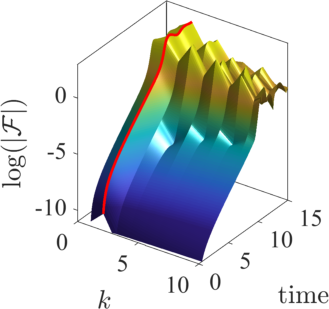}
    \hfill
    \includegraphics[width=0.22\textwidth]{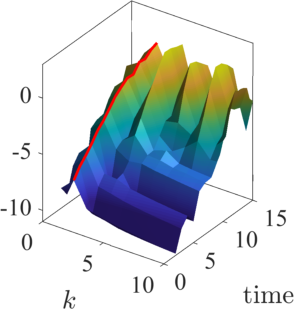}
    \hfill
    \includegraphics[width=0.22\textwidth]{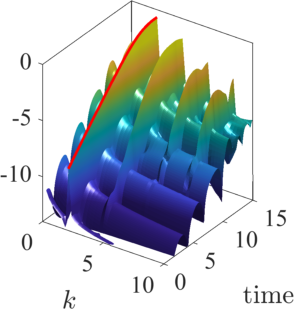}
    \hfill
    \includegraphics[width=0.18\textwidth]{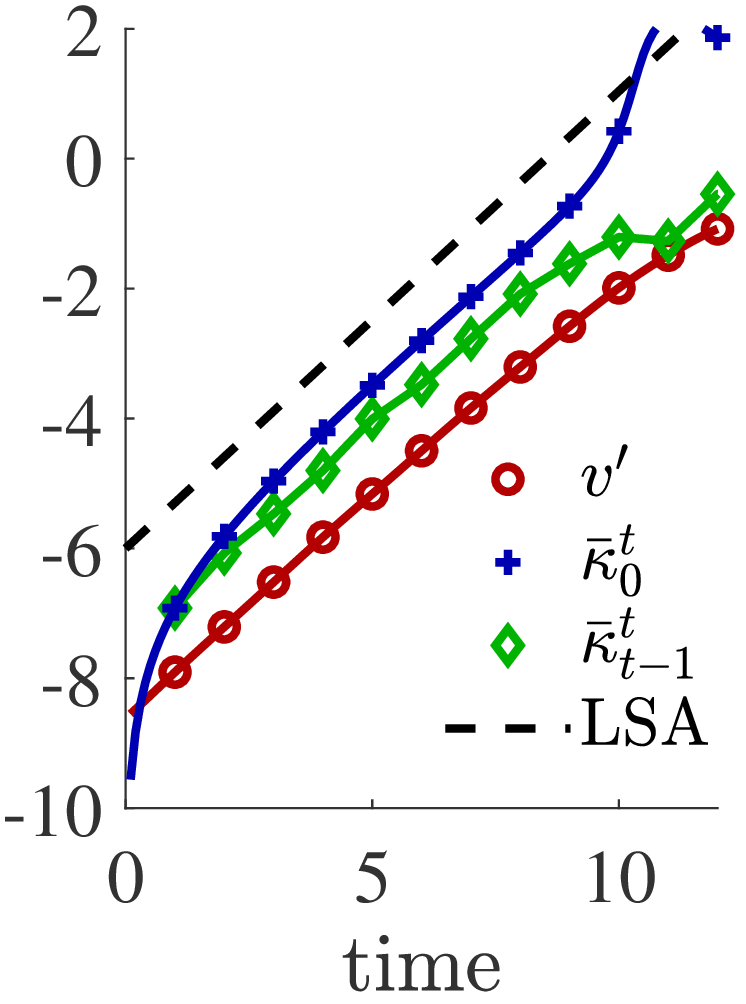}
    }
    \makebox[\textwidth][c]{
    \makebox[0.03\textwidth][l]{\rotatebox{90}{\hspace{20pt} $k_{pert}$ = $4.2$}}
	\hfill
    \includegraphics[width=0.25\textwidth]{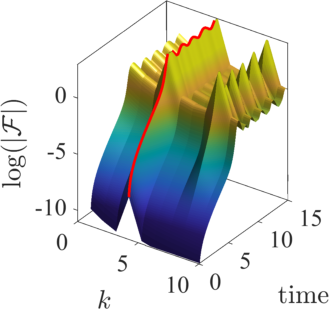}
    \hfill
    \includegraphics[width=0.22\textwidth]{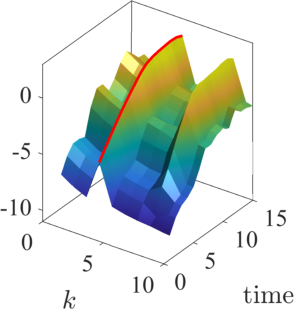}
    \hfill
    \includegraphics[width=0.22\textwidth]{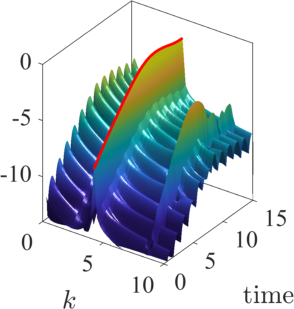}
    \hfill
    \includegraphics[width=0.18\textwidth]{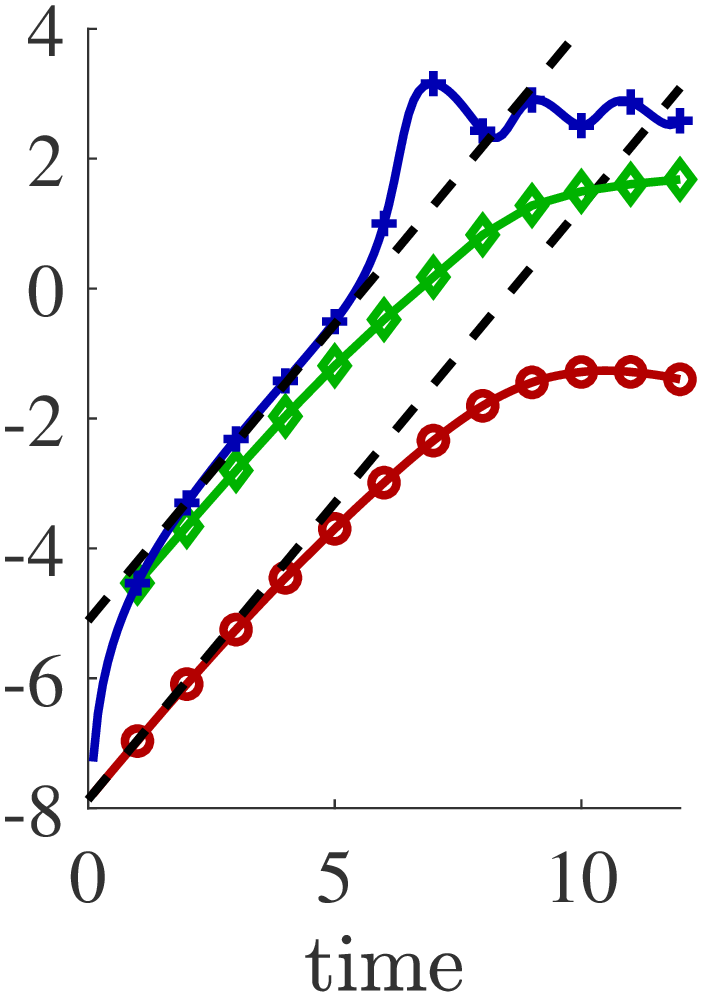}
    }
    \makebox[\textwidth][c]{
    \makebox[0.03\textwidth][l]{\rotatebox{90}{\hspace{20pt} $k_{pert}$ = $8.4$}}
	\hfill
    \includegraphics[width=0.25\textwidth]{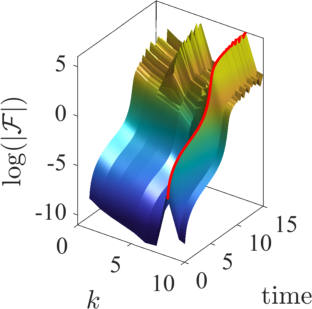}
    \hfill
    \includegraphics[width=0.22\textwidth]{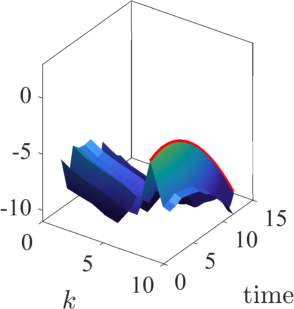}
    \hfill
    \includegraphics[width=0.22\textwidth]{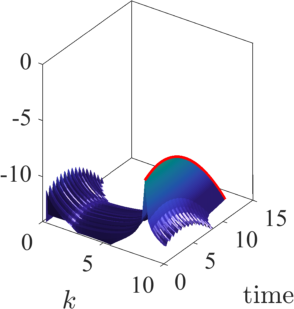}
    \hfill
    \includegraphics[width=0.18\textwidth]{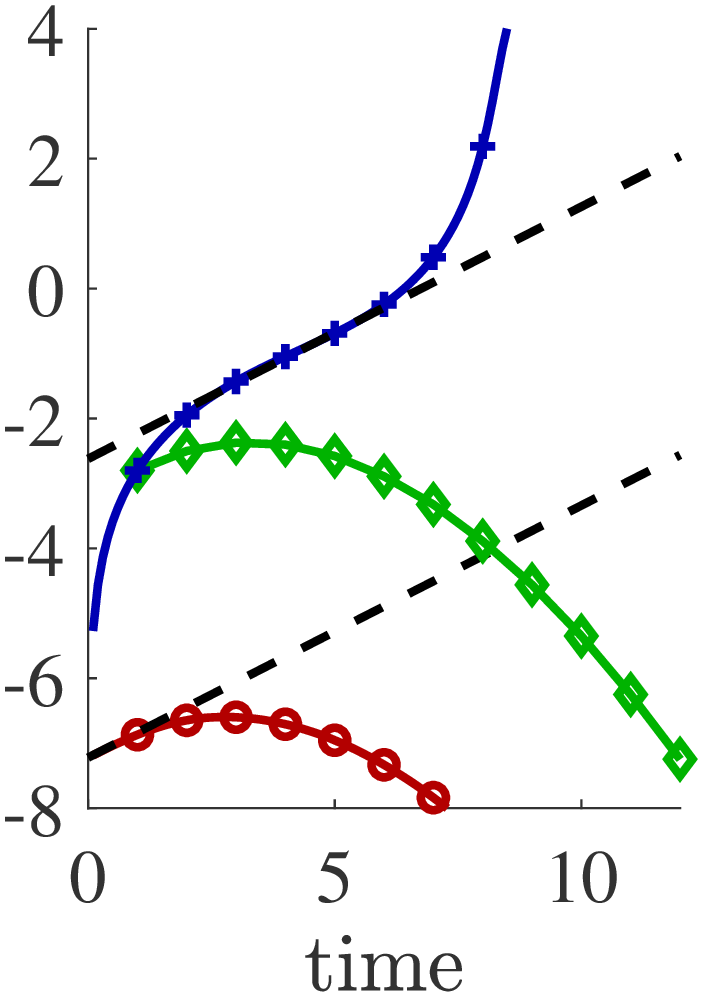}
    }
    \makebox[\textwidth][c]{
    \makebox[0.03\textwidth][l]{}
	\hfill
    \makebox[0.25\textwidth][c]{(a) $\bar{\kappa}_0^t$}
    \hfill
    \makebox[0.22\textwidth][c]{(b) $\bar{\kappa}_{t-1}^t$}
    \hfill
    \makebox[0.22\textwidth][c]{(c) $v'$}
    \hfill
    \makebox[0.18\textwidth][c]{(d) $\log(|\mathcal{F}(k=k_{pert})|)$}
    }
	\caption{
	Columns (a--c): amplitude of the Fourier transform $\mathcal{F}$ of $v'$ (a), $\bar{\kappa}_{0}^t$ (b), and  $\bar{\kappa}_{t-1}^t$ (c) plotted versus wavenumber and time. The temporal development of maxima in $\vert\mathcal{F}\vert$ at  is highlighted with a red line. 
	(d) Comparison of the amplitude of the Fourier transforms over time  with the LSA growth rate (dashed lines).
		The rows correspond to the wavenumbers of the initial perturbation mode with $k_{pert}$ = 2.1, 4.2, and 8.4.
    }
	\label{fig:temp_growth}
\end{figure}

To determine the growth of the maximum curvature change in the temporally developing jet flow, we consider the material line with a local maximum curvature gradient $\vert \mathbf{\nabla}\bar{\kappa}_0^8\vert$ and $\vert{\mathbf{\nabla}}\bar{\kappa}_0^6\vert$  according to the yellow markers in figures \ref{fig:alpha_2p1} and \ref{fig:alpha_4p2} respectively.
Along these lines, we evaluate the Fourier transform $\mathcal{F}(k)$ of $\bar{\kappa}_{0}^{t}$ and $\bar{\kappa}_{t-1}^{t}$ at every time step and record its amplitude, as shown in figure \ref{fig:temp_growth}(a--b) where surfaces of $|\mathcal{F}|$ are plotted over the wavenumber space and time.
Similarly, the $v'$ velocity amplitude is determined along the critical layer ($y/h$ = 0.5) and is added for reference in figure \ref{fig:temp_growth}(c).

All surface plots in figure \ref{fig:temp_growth} have a ridge of local maximum $|\mathcal{F}|$ at the respective perturbation wavenumber $k_{pert}$ (indicated as red line).
Additional ridges emerge over time in figure \ref{fig:temp_growth}(a--c). These are  an indication of  non-linear energy  transfer  to modes at other frequencies as the base profile diffuses over time. 

\begin{figure}
	\centering
    \includegraphics[width=0.223\textwidth]{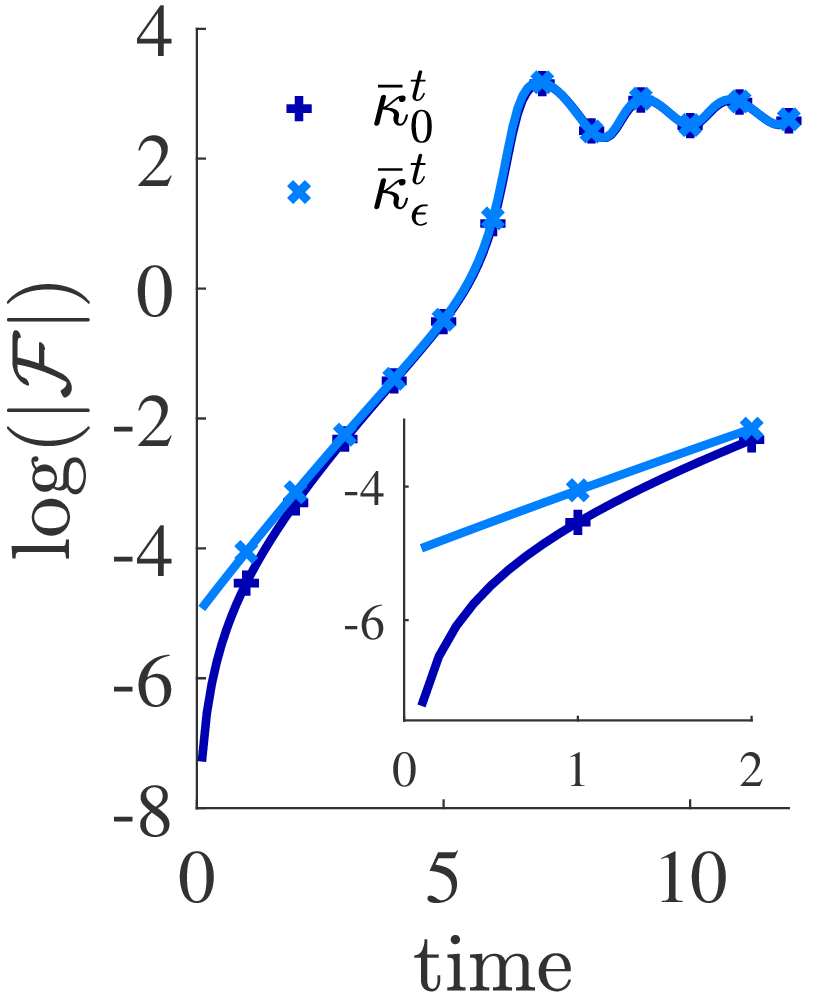}
	\caption{
	Comparison of the temporal development of the maximum amplitude of the Fourier transform $\mathcal{F}$ of $\bar{\kappa}_0^t$ and $\bar{\kappa}_\epsilon^t$ for $k_{pert}$ = 4.2, where $\epsilon$ is an initial perturbation of the material line.}
	\label{fig:temp_growth_pert}
\end{figure}

A direct comparison of of $v'$, $\bar{\kappa}_{0}^{t}$, and $\bar{\kappa}_{t-1}^{t}$ at the perturbation wavenumber is presented in figure \ref{fig:temp_growth}(d), where the  time-dependent Fourier transform amplitudes are plotted versus time. The slope of the dotted straight line, which is plotted for reference, is according to the exponential growth rate obtained from LSA, $e^{\omega_i(t_0) t}$.
The $v'$ velocity initially grows according to the rate predicted by LSA (dashed lines), but departs from the linear growth trend when the shear layer  diffuses significantly over time. 
While the $v'$ velocity growth starts from the imposed perturbation, the material lines have zero curvature change at the initial time. 
Because of this the growth of the material line's curvature change follows the growth predicted by LSA (dashed lines in figure \ref{fig:temp_growth}d) only after an initial adjustment phase.
This characteristic is confirmed by the analytic form of the curvature scalar equation \eqref{eq:kappa_barThm}.
An initial curvature can be added to the material line in the form of a sinuous perturbation (denoted by $\epsilon$) with the wavelength set to the LSA eigenmodes. 
In that case, the curvature change follows the growth predicted by LSA without the adjustment phase as shown in figure \ref{fig:temp_growth_pert}.

The Lagrangian curvature change for relatively short integration intervals,
$\bar{\kappa}_{t-1}^t$, (green) in figure \ref{fig:temp_growth}(d) closely follows
the trends of the $v'$ velocity (red). This is consistent with the  relation
between $\bar{\kappa}_{t-1}^t$ and $v'$ as derived in \eqref{eq:kdot}, as it can
be expected that the short time curvature change follows 
the infinitesimally small curvature change, $\dot{\kappa}_t$,
and thus also scales with the transverse velocity perturbation. 

\begin{figure}
    \centering
	\includegraphics[width=0.75\textwidth]{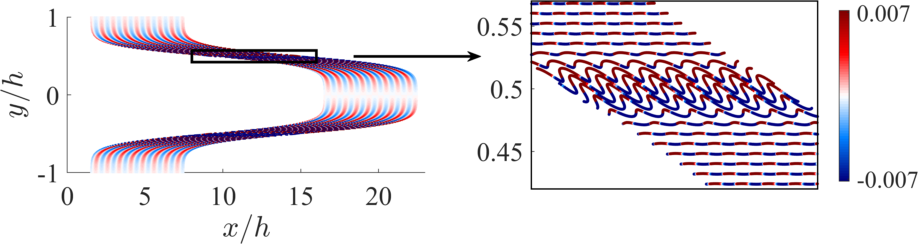}
	\caption{Curvature scalar field $\bar{\kappa}_{0}^{15.0}$  plotted along advected particle positions at $t$ = 15.0 for an initial perturbation mode with wavenumber $k_{pert}$ = 8.4. The right figure is a zoomed in visualization of the box identified in the left figure.}
	\label{fig:alpha_8p4}
\end{figure}

For larger time intervals, $T$, however, this reasoning  no longer applies because  non-linearities lead
to coupling with and excitation of other modes, and so \eqref{eq:kdot} which is derived for a single mode perturbation equation \eqref{eq:kappa_barThm} becomes invalid. For perturbations at higher wave number (figure \ref{fig:temp_growth}),
this mode coupling occurs at an earlier time as compared to cases with the lower wave number perturbations.
For example, in the case of perturbation with symmetric eigenmodes at $k_{pert}$ = 8.4 (see figure \ref{fig:temp_growth}d), 
$\bar{\kappa}_{0}^t$ grows whereas the $v'$ and $\bar{\kappa}_{t-1}^t$ reduce after a short initial
growth, i.e. the perturbation is ceasing and the flow becomes stable.
Owing to the low Reynolds number, the shear layer diffuses leading
to changes  in the stability properties over time and
as a result the mode at $k$ = 8.4 moves outside the unstable regime  (see figure \ref{fig:baseflow}b).
In the kinematic analysis, however, the initial perturbation of the fluid particles is carried on (unless the tracers are re-inialized) yielding a continuously growing deformation of the material lines within the layer 0.47 < $|y/h|$ < 0.53, as shown in figure \ref{fig:alpha_8p4}.
Notably, the curvature $\bar{\kappa}_{0}^t$ of the material line at $y/h$ = 0.5 follows the growth rate given by LSA (see figure \ref{fig:temp_growth}d).
This continuous growth of the deformation in the Lagrangian frame
showcases how the footprint of a temporary instability persists. 
This curvature persistence is a very useful
feature in practice as one necessarily has to identify instabilities over a finite time in experiments, while the
LSA and the instantaneous growth rate of $v'$ are generally unavailable.

\medskip

To asses the ability of the kinematic approach to identify emerging
unstable modes in a flow subject to  perturbations with multiple modes,
the jet flow is simulated with an initial random mode perturbation.
Following \citet{Kraichnan70}, the  velocity perturbation is initialized by superposition of fifty Fourier modes with random coefficients and amplitude as follows:
\begin{equation}
\label{eq:random}
    \mathbf{u}' = 2\sum_{n=1}^{N=50} \tilde{u}_{n}\boldsymbol{\sigma}_n \cos{\left(\mathbf{k}_n\cdot\mathbf{x} + \psi_n\right)}.
\end{equation}
Here, $\tilde{u}_n$ is the randomized amplitude function, $\mathbf{k}_n$ the wavenumber vector and $\psi_n$ a random phase shift.
By imposing a solenoidal velocity field, the condition $\mathbf{k}_n$ $\cdot$ $\boldsymbol{\sigma}_n$ = 0 determines the components of the unit vector $\boldsymbol{\sigma}$. 
Only multiples of the smallest possible wavenumber in the domain, i.e. $k_{x,y;n}$ $=$ $n k_{min}$, where $n$ is a random integer in the range 1 $\leq$ $n$ $\leq$ $N$, are considered to ensure periodicity.
The flow  is simulated at a higher Reynolds number of $Re_h$ = 300,000 
as compared to the cases above, which reduces the impact of the shear layer growth and the associated shifting of the stability properties of the flow (see figure \ref{fig:baseflow}b).

\begin{figure}
\centering 
	\includegraphics[width=0.9\textwidth]{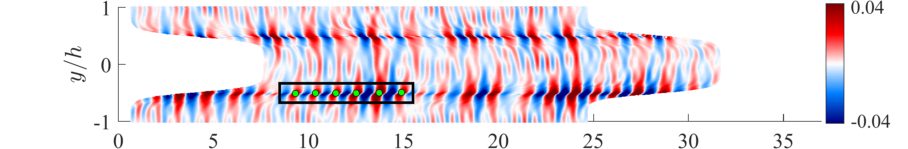}
	\makebox[\textwidth][c]{(a) $t$ = 7.0}
	\vskip\baselineskip
	\includegraphics[width=0.9\textwidth]{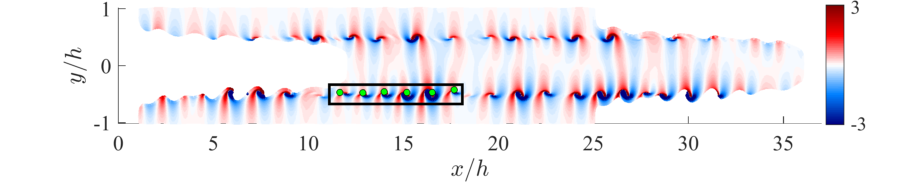}
	\makebox[\textwidth][c]{(b) $t$ = 11.0}
	\caption{Contours of the curvature scalar field $\bar{\kappa}_{0}^{7.0}$ (a) and $\bar{\kappa}_{0}^{11.0}$ (b) for perturbation with random modes.
	Green tracers: $\max(\bar{\kappa}_0^7)$ of the scalar field within the lower half of the jet shown in (a) and advected positions at $t$ = 11 in (b).
	}
	\label{fig:alpha_rand}
\end{figure}

Snapshots of the particle trace colored by the curvature $\bar{\kappa}_0^t$ are presented in figure \ref{fig:alpha_rand} for integration times $T$ = 7 (a) and $T$ = 11 (b).
The random modes  leave a footprint in the particle trace 
and high frequency content is still visible at $t$ = 7.
At this time, however, the dominant modes driving the fluid deformation have emerged in the curvature field 
and are decisive in the the topology of the material lines at $t$ = 11 (see figure \ref{fig:alpha_rand}b).
To understand the connection between the curvature field
at $t$ = 7 and $t$ = 11, fluid tracers are initialized at local curvature maxima at $t$ = 7 (see green markers).
At $t$ = 11, the material lines have deformed significantly along the shear layers, leading
to increased local magnitudes of the maximum curvature. 
Figure \ref{fig:alpha_rand}(b) shows that the green tracers initialized at $t$ = 7 are located at the maximum fluid wrinkling and curvature locations of the forming vortices at $t$ = 11, underscoring that local curvature maxima can identify locations of vortex roll-up at early times.

\subsection*{NACA 65(1)-412 airfoil flow}
\begin{figure}
    \centering
    \includegraphics[width=0.7\textwidth]{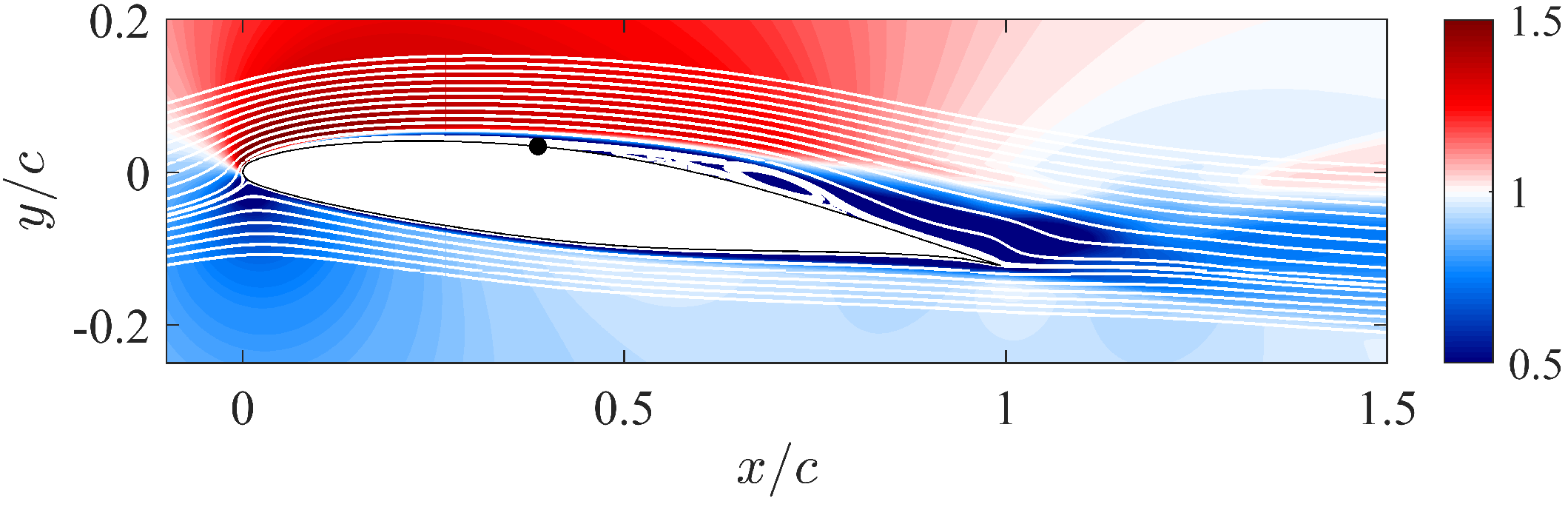}
	\caption{Contours of time-averaged velocity magnitude and streamlines (white) for a two-dimensional Navier-Stokes
	flow over the NACA-65(1)412 airfoil at 7$^\circ$ incidence.
	The black circle identifies the time-averaged zero skin friction point.
	Plot shows rotated field with horizontally aligned inflow velocity vector.
	}
	\label{fig:airfoil_avg}
\end{figure}

The transitional flow over a NACA 65(1)-412 airfoil at 7$^\circ$ incidence and chord-based Reynolds number $Re_c$ = 20,000 is characterized by  a laminar separation bubble (LSB) as shown by the time averaged velocity field and streamlines
in figure \ref{fig:airfoil_avg}.
Upstream of the separation bubble, fluid particles
well up as discussed in the introduction and depicted in figure \ref{fig:material_line}(e-g). They
consequently move towards an asymptotic manifold that has its origin at the averaged zero  skin friction point (figure \ref{fig:airfoil_avg}) 
where the streamlines break away from the wall.

\begin{figure}
	\makebox[\textwidth][c]{
    \includegraphics[width=0.481\textwidth]{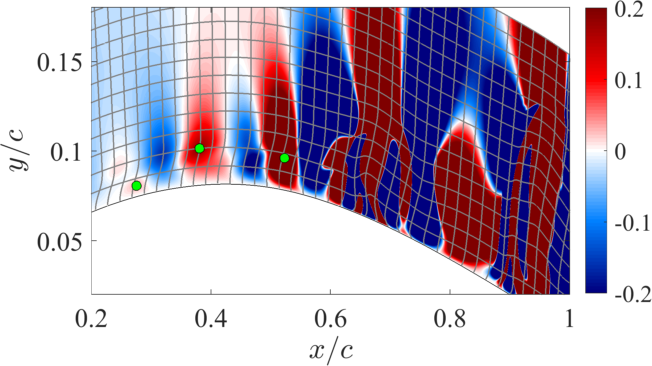}
    \hfill
    \includegraphics[width=0.44\textwidth]{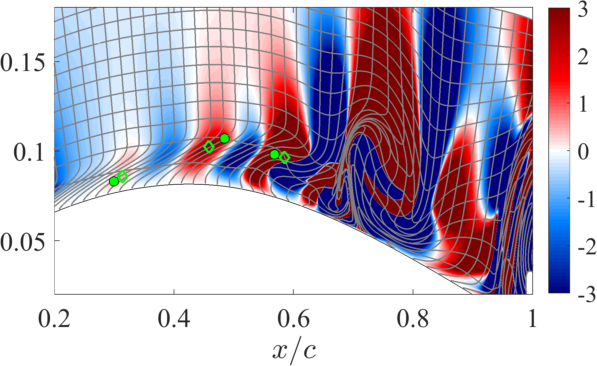}
    }
    \makebox[\textwidth][c]{
    \makebox[0.45\textwidth][c]{(a) $\bar{\kappa}_{55.0}^{55.01}$}
    \hfill
    \makebox[0.45\textwidth][c]{(b) $\bar{\kappa}_{55.0}^{55.1}$}
    }
	\caption{Contour plots of the curvature scalar $\bar{\kappa}_{55.0}^{55.01}$ (a) and $\bar{\kappa}_{55.0}^{55.1}$ (b) with advected Lagrangian grid in gray.
	Local maxima of the curvature scalar in green: 
	$\max(\bar{\kappa}_{55.0}^{55.01})$ (circles) are based on the scalar field in (a) and their advected positions plotted in (b). $\max(\bar{\kappa}_{55.0}^{55.1})$ (diamonds) are based on the scalar field in (b).
	The $y$-axis is stretched.}
	\label{fig:airfoil_kappa}
\end{figure}

Upon flow separation, a shear layer forms that  is unstable as  plots of the instantaneous Lagrangian curvature change field $\bar{\kappa}_{55.0}^{55.01}$ and $\bar{\kappa}_{55.0}^{55.1}$ (based on wall-parallel material lines) in figure \ref{fig:airfoil_kappa}(a--b) show.
In the figure, the $y$ coordinate is independently scaled from the $x$-coordinate to obtain a clearer visualization of the flow
phenomena that occur in a thin region close to the airfoil.
Kelvin-Helmholtz instabilities along the spatially
developing, separated  shear layer lead to vortex
formation upstream of the location where the LSB reattaches. Further downstream, the flow transitions to a wall-bounded, vortex dominated pattern.
The curvature fields $\bar{\kappa}_{55.0}^{55.01}$ and $\bar{\kappa}_{55.0}^{55.1}$ show the wrinkling modes associated with the Kelvin-Helmholtz instabilities. These modes are  visible through bands
of positive and negative curvatures in the vicinity the separated shear layer similar to the temporally developing jet.

The bands in the Lagrangian curvature fields in figure \ref{fig:airfoil_kappa}(a) and (b)  upstream of the vortex roll-up at $x/c$ = 0.7 identify
the process of off-wall material line folding associated with the Kelvin-Helmholtz instability in the shear layer.
The green, circular markers identify the  locations of the local  maxima of $\bar{\kappa}_{55.0}^{55.01}$ on these curvature bands in figure \ref{fig:airfoil_kappa}(a) and their advected locations at $t$ = 55.1  in figure \ref{fig:airfoil_kappa}(b).
The advected Lagrangian grids (grey) based on the tracers initialized according to figure \ref{fig:airf_cyl_mesh}(a) are also plotted for reference.
Similar to the jet flow, the maximum wrinkling occurs in the vicinity of the curvature extrema. These curvature peaks are 
also nearly material and remain close to the local maxima of the curvature computed over a longer integration interval, $\bar{\kappa}_{55.0}^{55.1}$ (see green, diamond-shaped markers).
The strongly increasing magnitude of the curvature maxima along the spatially developing unstable shear flow
provides evidence that curvature changes identify instabilities and
the material wrinkling at the onset of vortex formation.

\subsection*{Wake behind a circular cylinder}

\begin{figure}
	\makebox[\textwidth][c]{
	\includegraphics[height=0.137\textwidth]{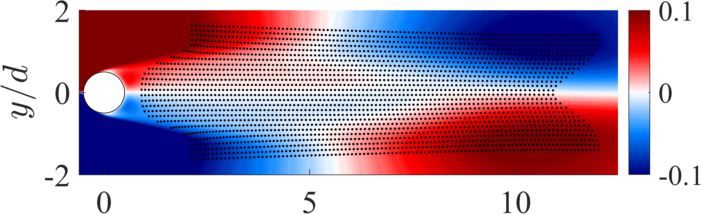}
	\hfill
	\includegraphics[height=0.137\textwidth]{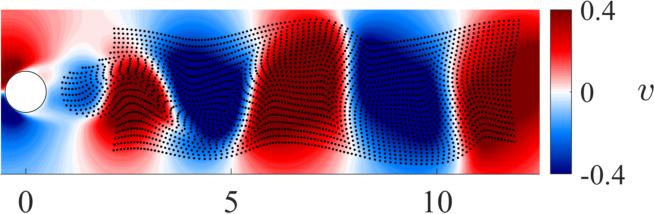}
	}
	\vskip\baselineskip
	\makebox[\textwidth][c]{
	\includegraphics[height=0.137\textwidth]{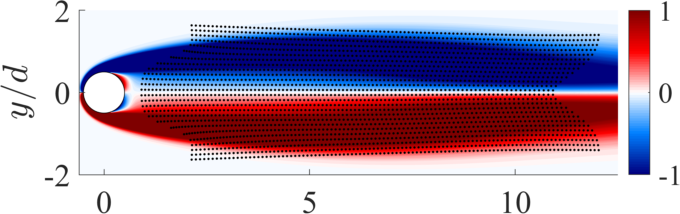}
	\hfill
	\includegraphics[height=0.137\textwidth]{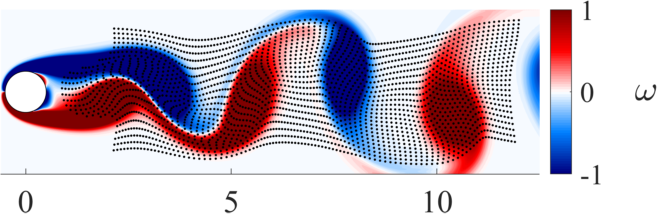}
	}
	\vskip\baselineskip
	\makebox[\textwidth][c]{
	\includegraphics[height=0.137\textwidth]{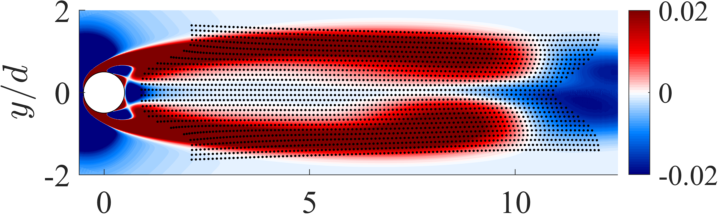}
	\hfill
	\includegraphics[height=0.137\textwidth]{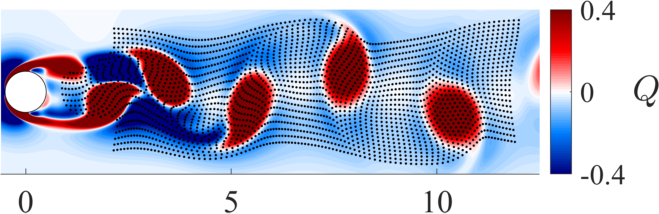}
	}
	\vskip\baselineskip
	\makebox[\textwidth][c]{
	\includegraphics[height=0.137\textwidth]{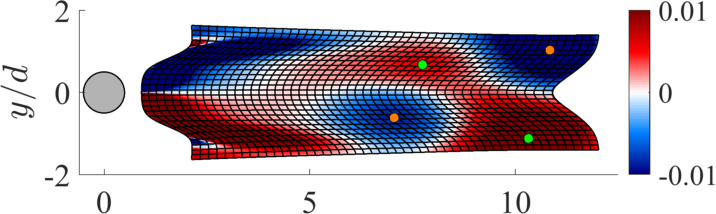}
	\hfill
	\includegraphics[height=0.137\textwidth]{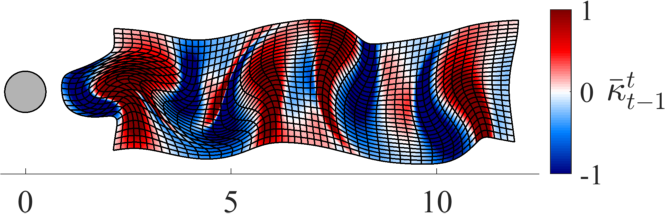}
	}
	\vskip\baselineskip
	\makebox[\textwidth][c]{
	\includegraphics[height=0.166\textwidth]{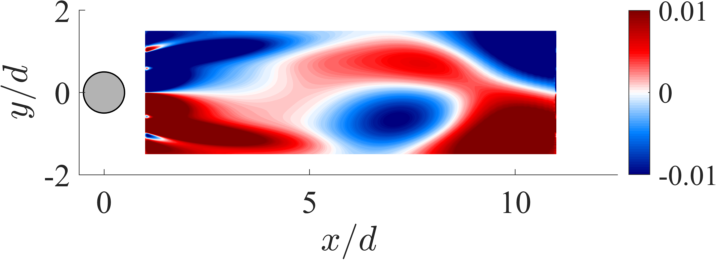}
	\hfill
	\includegraphics[height=0.166\textwidth]{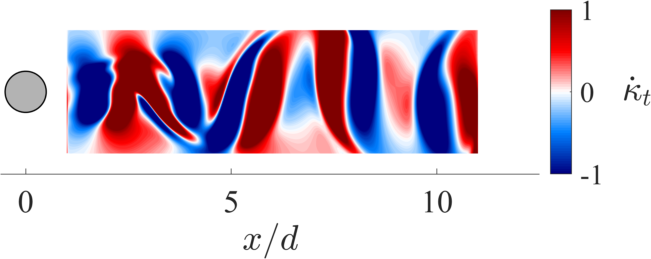}
	}
	\makebox[\textwidth][c]{
	\makebox[0.45\textwidth][c]{(a) $t$ = 105}
    \hfill
    \makebox[0.45\textwidth][c]{(b) $t$ = 160}
    }
	\caption{Comparison of Eulerian and Lagrangian quantities for the circular cylinder. From top to bottom: $v$ velocity, vorticity, Q-criterion, Lagrangian curvature scalar field $\bar{\kappa}_{t-1}^{t}$ (with deformed Lagrangian grid) and Lagrangian curvature rate $\dot{\kappa}_{t}$ at times $t$ = 105 (a) and $t$ = 160 (b). 
	Local minima and maxima of $\bar{\kappa}_{104}^{105}$ indicated by orange and green markers in (a).
	The particle trace is added to the plots of Eulerian quantities for orientation. Note that the range of the color map changes with $t$.}
	\label{fig:cyl_kappa}
\end{figure}

To understand the material wrinkling and its curvature field related to the emergence of an unstable mode in the growing wake of a circular cylinder, horizontal material lines are initialized half a diameter downstream of the cylinder and advected for one convective time unit. 
The Lagrangian curvature change ($\bar{\kappa}_{t-1}^{t}$) and curvature rate ($\dot{\kappa}_{t}$) are extracted from the material lines and contours fields
thereof are plotted in figure \ref{fig:cyl_kappa} at times $t$ = 105 and $t$ = 160, together with common Eulerian quantities such as the transverse velocity component, vorticity and the Q-criterion.  Also visualized for reference
are the advected and deformed Lagrangian grids, which are overlayed to the $\bar{\kappa}_{t-1}^{t}$ fields. 

Up to $t$ = 105, the recirculation region is growing with time.
No steady time-averaged solution is hence available that would allow for the computation of perturbation quantities. 
While all (frame-dependent) Eulerian quantities in figure \ref{fig:cyl_kappa} show a symmetric   topology with respect to the x axis at $t$ = 105, the curvature scalars $\bar{\kappa}_{104}^{105}$ and $\dot{\kappa}_{105}$ show an asymmetric fluid deformation at this early stage. 
This asymmetry is the early indicator of the development of 
the wake instability that remains hidden to the other fields.

At a later time ($t$ = 160), the wake is unstable and has developed the characteristic Von-Karman vortex street, where the strongly increased transverse fluid motion results in significant bending of the material lines visualized through the deformed material grid along with $\bar{\kappa}_{159}^{160}$.
While vortices in the wake are identified by regions of large vorticity or positive values of the Q-criterion, their rotating motion induces pairs of negative and positive curvature that resemble the \textit{yin-yang} symbol in the $\bar{\kappa}_{159}^{160}$ field.
Connectors between vortices are identified as slender curvature ridges and positive and negative values provide directional information about the fluid motion and deformations.

\section{Conclusion} \label{conclusion}
A finite-time curvature change diagnostic is introduced to identify 
instabilities in material lines in the Lagrangian frame. 
By defining a flow instability in the Lagrangian frame as the increased folding of lines of fluid particles, the identification does not require knowledge of the base flow profile or averaged flow fields but only the flow map of particle traces.
Material lines are sensitive to fluid deformation and can therefore locate unstable modes in the particle phase early on.
Because the transient short-term instabilities leave a footprint in the advected fluid material 
line's curvature change over finite time, a convenient time interval would
be available in practice to capture instabilities. 
The material curvature is objective, independent of the parametrization, and so the identification
applies to a broad variety of flows, including  those over general complex geometries, rotating frames, and experimental setups with particle tracers. 

\textcolor{black}{Analytic formulas for the approximation of the flow map and curvature change are provided for perturbed parallel shear flows. These formulas connect relevant Eulerian quantities used in LSA characterizing the dynamics of perturbations in the infinite-dimensional space of velocities with their induced effects in the physical space of fluid flows.}
The workings of the diagnostic are illustrated conceptually for the case of the hyperbolic tangent jet in a periodic domain for perturbations with eigenmodes of three different wavenumbers and random noise.
It is shown that the material curvature change accurately captures the unstable modes early in the jet flow. Exponential growth rates of Lagrangian curvature match the values predicted by linear stability analysis, as we predict analytically. In a jet flow with random initial mode perturbations, the curvature field  captures the emerging unstable mode.

Tests of the flows over a cambered airfoil at incidence and the wake behind a circular cylinder demonstrate the ability of the kinematic approach to identify instabilities without the requirement of an analytical mean or a frame of reference.
The curvature field  identifies the early onset of vortex formation 
that remains hidden to typical Eulerian diagnostics.\\

\section*{Acknowledgments}
We gratefully acknowledge funding provided by the Unsteady Aerodynamics, and Turbulent Flow program and the Computational Mathematics program of the Air Force Office of Scientific Research under grants FA9550-16-1-0392and FA9550-19-1-0387, respectively, and from Solar Turbines. M.S. acknowledges support from the Schmidt Science Fellowship and the Postdoc Mobility Fellowship from the Swiss National Foundation. Declaration of Interests: The authors report no conflict of interest.

\appendix
\section{Proof of \texorpdfstring{\boldmath{$\dot{\kappa}_{t}$} = \boldmath{$-\partial_{xx}v$}}{kdot}} \label{appendix_kappa}
Following \citet{serra18}, the material curvature rate of material lines parametrized by arc-length is
\begin{equation}\label{eq:kt}
\dot{\kappa}_{t} = \langle(\nabla\mathbf{S}\mathbf{r}')\mathbf{r}',\mathbf{R}\mathbf{r}'\rangle
-\frac{1}{2}\langle\nabla\omega,\mathbf{r}'\rangle
-3\kappa_0\langle\mathbf{r}',\mathbf{S}\mathbf{r}'\rangle,
\end{equation}
with the rate-of-strain tensor $\mathbf{S}$ = $\frac{1}{2}(\nabla\mathbf{u}+\nabla\mathbf{u}^\top)$, the local tangent vector $\mathbf{r}', \vert \mathbf{r}'\vert =1$, and $\mathbf{R}$ is a rotation matrix defined according to \eqref{eq:R}.

We now separate the right-hand side of \eqref{eq:kt} into 3 parts and simplify each term separately under the assumption $\mathbf{r}'$ = $[1, 0]^\top$.

\vskip\baselineskip
\textit{Evaluation of the first term:} $\langle(\nabla\mathbf{S}\mathbf{r}')\mathbf{r}',\mathbf{R}\mathbf{r}'\rangle$

\begin{equation}
\begin{split}
\nabla\mathbf{S}\mathbf{r}' &= \partial_x\mathbf{S}r_x' + \partial_y\mathbf{S}r_y' \\
&= \partial_x
\begin{bmatrix}
\partial_x u & (\partial_y u+\partial_x v)/2 \\ (\partial_x v+\partial_y u)/2 & \partial_y v
\end{bmatrix}r_x'
+ \partial_y
\begin{bmatrix}
\partial_x u & (\partial_y u+\partial_x v)/2 \\ (\partial_x v+\partial_y u)/2 & \partial_y v
\end{bmatrix}r_y' \\
&= \begin{bmatrix}
\partial_{xx} u r_x' & r_x'(\partial_{yx} u+\partial_{xx} v)/2 \\ r_x'(\partial_{xx} v+\partial_{yx} u)/2 & \partial_{yx} v r_x'
\end{bmatrix}
+ \begin{bmatrix}
\partial_{xy} u r_y' & r_y'(\partial_{yy} u+\partial_{xy} v)/2 \\ r_y'(\partial_{xy} v+\partial_{yy} u)/2 & \partial_{yy} v r_y'
\end{bmatrix}
\end{split}
\end{equation}

\begin{equation}
\begin{split}
(\nabla\mathbf{S}\mathbf{r}')\mathbf{r}' &= 
\begin{bmatrix}
\partial_{xx} u r_x'^2 + r_x'r_y'(\partial_{yx} u+\partial_{xx} v)/2 \\ r_x'^2(\partial_{xx} v+\partial_{yx} u)/2 + \partial_{yx} v r_x'r_y'
\end{bmatrix}
+ \begin{bmatrix}
\partial_{xy} u r_y'r_x' + r_y'^2(\partial_{yy} u+\partial_{xy} v)/2 \\ r_y'r_x'(\partial_{xy} v+\partial_{yy} u)/2 + \partial_{yy} v r_y'^2
\end{bmatrix} \\
&= \begin{bmatrix}
\partial_{xx} u r_x'^2 + r_x'r_y'(\partial_{yx} u+\partial_{xx} v)/2 + \partial_{xy} u r_y'r_x' + r_y'^2(\partial_{yy} u+\partial_{xy} v)/2 \\ r_x'^2(\partial_{xx} v+\partial_{yx} u)/2 + \partial_{yx} v r_x'r_y' + r_y'r_x'(\partial_{xy} v+\partial_{yy} u)/2 + \partial_{yy} v r_y'^2
\end{bmatrix}
\end{split}
\end{equation}

\begin{equation}\label{eq:app_a_1}
\begin{split}
\langle(\nabla\mathbf{S}\mathbf{r}')\mathbf{r}',\mathbf{R}\mathbf{r}'\rangle & =
\partial_{xx} u r_x'^2r_y' + r_x'r_y'^2(\partial_{yx} u+\partial_{xx} v)/2 + \partial_{xy} u r_y'^2r_x' + r_y'^3(\partial_{yy} u+\partial_{xy} v)/2 \\
&- r_x'^3(\partial_{xx} v+\partial_{yx} u)/2 - \partial_{yx} v r_x'^2r_y' - r_y'r_x'^2(\partial_{xy} v+\partial_{yy} u)/2 - \partial_{yy} v r_y'^2r_x'
\end{split}
\end{equation}

For a tangent vector $\mathbf{r}'$ = $[1, 0]^\top$, \eqref{eq:app_a_1} can be rewritten as
\begin{equation}
\langle(\nabla\mathbf{S}\mathbf{r}')\mathbf{r}',\mathbf{R}\mathbf{r}'\rangle = -(\partial_{xx} v+\partial_{yx} u)/2.
\end{equation}

\vskip\baselineskip
\textit{Evaluation of the second term:} $-\frac{1}{2}\langle\nabla\omega,\mathbf{r}'\rangle$

\begin{equation}\label{eq:app_a_2}
-\frac{1}{2}\langle\nabla\omega,\mathbf{r}'\rangle = -(\partial_x\omega r_x' + \partial_y\omega r_y')/2
\end{equation}

Again, we assume a tangent vector $\mathbf{r}'$ = $[1, 0]^\top$, which simplifies \eqref{eq:app_a_2} to 
\begin{equation}
-\frac{1}{2}\langle\nabla\omega,\mathbf{r}'\rangle = -\partial_x\omega/2.
\end{equation}

\vskip\baselineskip
\textit{Evaluation of the third term:} $-3\kappa_0\langle\mathbf{r}',\mathbf{S}\mathbf{r}'\rangle$

\begin{equation}
\mathbf{S}\mathbf{r}' = \begin{bmatrix}
\partial_x u r_x' + r_y'(\partial_y u+\partial_x v)/2 \\ r_x'(\partial_x v+\partial_y u)/2 + \partial_y v r_y'
\end{bmatrix}
\end{equation}

\begin{equation}
-3\kappa_0\langle\mathbf{r}',\mathbf{S}\mathbf{r}'\rangle = 
-3\kappa_0(\partial_x u r_x'^2 + r_y'r_x'(\partial_y u+\partial_x v)/2 + r_x'r_y'(\partial_x v+\partial_y u)/2 + \partial_y v r_y'^2)
\end{equation}

Again, simplification with $\mathbf{r}'$ = $[1, 0]^\top$ leads to
\begin{equation}
-3\kappa_0\langle\mathbf{r}',\mathbf{S}\mathbf{r}'\rangle = 
-3\kappa_0\partial_x u,
\end{equation}
where the initial curvature $\kappa_0$ is also zero so that
\begin{equation}
-3\kappa_0\langle\mathbf{r}',\mathbf{S}\mathbf{r}'\rangle = 0.
\end{equation}

\vskip\baselineskip
Substituting the three simplified terms back into \eqref{eq:kt} yields
\begin{equation}
\dot{\kappa}_{t} = \langle(\nabla\mathbf{S}\mathbf{r}')\mathbf{r}',\mathbf{R}\mathbf{r}'\rangle
-\frac{1}{2}\langle\nabla\omega,\mathbf{r}'\rangle
-3\kappa_0\langle\mathbf{r}',\mathbf{S}\mathbf{r}'\rangle
= -\frac{1}{2}(\partial_{xx} v+\partial_{yx} u + \partial_x\omega),
\end{equation}
and with $\omega$ = $\partial_x v - \partial_y u$ it follows that
\begin{equation}
\dot{\kappa}_{t} = -\frac{1}{2}\partial_x(\partial_{x} v+\partial_{y} u + \omega)
= -\frac{1}{2}\partial_x(\partial_{x} v+\partial_{y} u + \partial_x v -\partial_y u)
= -\partial_{xx}v.
\end{equation}
\section{Proof of Theorem \ref{sec:Thm1}} \label{appendix_kbar}
Here we derive an analytic expression for the curvature change $\bar{\kappa}_{t_{0}}^{t_0+T}$ in perturbed parallel shear flows of the form
\begin{equation}\label{eq:uepsilon}
\mathbf{u_{\epsilon}} = \mathbf{u_0}+\epsilon\mathbf{u^\prime}, \quad 0<\epsilon\ll 1,
\end{equation}
where $\mathbf{u_0}=[U(y), 0]^\top$ is the base flow and $\mathbf{u^\prime}=\operatorname{Re}([\hat{u}(y),\hat{v}(y)]e^{i(kx - \omega t)})$ is the perturbation, whose components are  
\begin{equation}
\begin{split}
    u' &= e^{\omega_i t}\left[\hat{u}_r\cos(kx-\omega_r t)-\hat{u}_i\sin(kx-\omega_r t)\right], \\
    v' &= e^{\omega_i t}\left[\hat{v}_r\cos(kx-\omega_r t)-\hat{v}_i\sin(kx-\omega_r t)\right].
    \end{split}
\end{equation}
Here, $\hat{u}=\hat{u}_r + i \hat{u}_i,\ \hat{v}=\hat{v}_r + i \hat{v}_i$ are complex amplitudes,  $\omega=\omega_r + i\omega_i$ is the complex frequency and $k$ the real wave number.
Because analytic solutions of of the flow map $\mathbf{{_\epsilon}F}$ induced by $\mathbf{u_{\epsilon}}$ are not available, we first seek an analytic approximation of $\mathbf{{_\epsilon}F}$ by neglecting $\mathcal{O}(\epsilon^2)$ terms. We define the flow map of the base flow and the perturbed flows as 
\begin{equation}
\begin{split}
    \mathbf{{{_0}F}}_{t_0}^{t_0+T}(\mathbf{x_0}) &= \mathbf{x}_0 + \int_{t_0}^{t_0+T}\mathbf{u_0}\left(\mathbf{{{_0}F}}_{t_0}^{\tau}(\mathbf{x_0}),\tau\right)\mathrm{d}\tau, \\
    \mathbf{{{_\epsilon}F}}_{t_0}^{t_0+T}(\mathbf{x_0}) &= \mathbf{x}_0 + \int_{t_0}^{t_0+T}\mathbf{u_\epsilon}\left(\mathbf{{{_\epsilon}F}}_{t_0}^{\tau}(\mathbf{x_0}),\tau\right)\mathrm{d}\tau.
    \end{split}
    \label{eq:Perturb_base_F}
\end{equation}
Computing a Taylor expansion of $\mathbf{{_\epsilon}F}$ with respect to $\epsilon$, we obtain 
\begin{equation}
    \mathbf{{{_\epsilon}F}}_{t_0}^{t_0+T}(\mathbf{x_0}) - \mathbf{{{_0}F}}_{t_0}^{t_0+T}(\mathbf{x_0}) = \underbrace{\frac{d}{d\epsilon} \mathbf{{{_\epsilon}F}}_{t_0}^{t_0+T}(\mathbf{x_0})\vert_{\epsilon=0}}_{\mathbf{A}_{t_{0}}^{t_{0}+T}(\mathbf{x_0})}\epsilon + \mathcal{O}(\epsilon^2),
     \label{eq:TaylorExpF}
\end{equation}
where $\mathbf{A}_{t_{0}}^{t_{0}+T}(\mathbf{x_0})$ is a vector governing the leading order deviations between trajectories of the base flow and the perturbed flow. From equations (\ref{eq:Perturb_base_F}-\ref{eq:TaylorExpF}), we find that $\mathbf{A}_{t_{0}}^{t_{0}+T}(\mathbf{x_0})$ satisfies the linear non-autonomous vectorial differential equation 
\begin{equation}
	\begin{cases}
		\dot{\overline{\mathbf{A}_{t_{0}}^{t}}}(\mathbf{x_0})=\mathbf{\nabla u_0}(\mathbf{{{_0}F}}_{t_0}^t(\mathbf{x_0}))\mathbf{A}_{t_{0}}^{t}(\mathbf{x_0}) + \mathbf{{u^\prime}}(\mathbf{{{_0}F}}_{t_0}^{t}(\mathbf{x_0}),t)\\
		\mathbf{A}_{t_{0}}^{t_0}(\mathbf{x_0}) = \mathbf{0}\\
		\mathbf{{{_0}F}}_{t_0}^{t}(\mathbf{x_0}) = \mathbf{x_0} + \mathbf{u_0}(\mathbf{x_0})(t-t_0).
	\end{cases}
	\label{eq:ADiffeq_b}
\end{equation}
The dynamic matrix of equation (\ref{eq:ADiffeq_b}) is the Jacobian of the base flow, while the forcing term is the perturbation, both of which evaluated along trajectories of the base flow. These terms as well as (\ref{eq:ADiffeq_b}) admit analytic solutions, hence providing the following analytic leading order approximation of the perturbed flow map
\begin{align}\label{eq:flowmap_eps}
    \mathbf{{{_\epsilon}F}}_{t_0}^{t_0+T}(\mathbf{x_0}) &= \mathbf{{{_0}F}}_{t_0}^{t_0+T}(\mathbf{x_0}) + \mathbf{A}_{t_{0}}^{t_0+T}(\mathbf{x_0})\epsilon + \mathcal{O}(\epsilon^2)\nonumber\\
    &=\begin{bmatrix}
    x_0 + U(y_0)T + \epsilon\left(f_1(y_0)\hat{u}_r(y_0) + f_2(y_0)\hat{u}_i(y_0)\right)/g(y_0) \\
    y_0 + \epsilon\left(f_5(y_0)\hat{v}_r(y_0) + f_6(y_0)\hat{v}_i(y_0)\right)/g(y_0)
    \end{bmatrix} \\
    &+\begin{bmatrix}
    \epsilon\left(f_3(y_0)\hat{v}_r(y_0)\:\partial_y U(y_0) + 
    f_4(y_0)\hat{v}_i(y_0)\:\partial_y U(y_0)\right)/g^2(y_0) \\
    0
    \end{bmatrix} + \mathcal{O}(\epsilon^2), \nonumber
\end{align}
where
\begin{equation}
\begin{split}
    f_1 &= -e^{\omega_i t_0}\omega_i \cos(k x_0-\omega_r t_0) + e^{\omega_i t_0}\Big(\omega_r-k U(y_0)\Big) \sin(k x_0-\omega_r t_0) \\
        &+e^{\omega_i(t_0+T)} \omega_i \cos(k x_0-\omega_r (t_0+T)+k U(y_0) T) \\
        &-e^{\omega_i(t_0+T)} \Big(\omega_r-k U(y_0)\Big) \sin(k x_0-\omega_r (t_0+T)+k U(y_0) T),
\end{split}
\end{equation}
\begin{equation}
\begin{split}
    f_2 &= e^{\omega_i t_0}\omega_i \sin(k x_0-\omega_r t_0)+e^{\omega_i t_0}\Big(\omega_r-k U(y_0)\Big) \cos(k x_0-\omega_r t_0) \\
        &-e^{\omega_i(t_0+T)} \omega_i \sin(k x_0-\omega_r (t_0+T)+k U(y_0) T) \\
        &-e^{\omega_i(t_0+T)} \Big(\omega_r-k U(y_0)\Big)\cos(k x_0-\omega_r (t_0+T)+k U(y_0) T),
\end{split}
\end{equation}
\begin{equation}
\begin{split}
    f_3 &= -e^{\omega_i t_0}\Big(\omega_i^2+\omega_i^3 T-\omega_r^2+\omega_i\omega_r^2 T-k^2     U^2(y_0) + \omega_i k^2 U^2(y_0) T \\
        &+ 2 \omega_r k U(y_0) -2 \omega_i \omega_r k U(y_0) T\Big) \cos(k x_0-\omega_r t_0) \\
        &+e^{\omega_i t_0}\Big(2 \omega_i \omega_r +\omega_i^2 \omega_r T+\omega_r^3 T -k^3 U^3(y_0) T +3 \omega_r k^2 U^2(y_0) T - 2\omega_i k U(y_0) \\
        &- \omega_i^2 k U(y_0) T - 3 \omega_r^2 k U(y_0) T \Big) \sin(k x_0-\omega_r t_0) \\
        &+e^{\omega_i(t_0+T)} \Big(\omega_i^2-\omega_r^2-k^2 U^2(y_0) + 2 \omega_r k U(y_0)\Big) \cos(k x_0-\omega_r (t_0+T)+k U(y_0) T) \\
        &-e^{\omega_i(t_0+T)} \Big(2 \omega_i \omega_r - 2 \omega_i k U(y_0)\Big) \sin(k x_0-\omega_r (t_0+T)+k U(y_0) T),
\end{split}
\end{equation}
\begin{equation}
\begin{split}
    f_4 &= e^{\omega_i t_0}\Big(\omega_i^2+\omega_i^3 T-\omega_r^2+\omega_i\omega_r^2 T-k^2     U^2(y_0) + \omega_i k^2 U^2(y_0) T \\
        &+ 2 \omega_r k U(y_0) -2 \omega_i \omega_r k U(y_0) T\Big) \sin(k x_0-\omega_r t_0) \\
        &+e^{\omega_i t_0}\Big(2 \omega_i \omega_r +\omega_i^2 \omega_r T+\omega_r^3 T -k^3 U^3(y_0) T +3 \omega_r k^2 U^2(y_0) T - 2\omega_i k U(y_0) \\
        &- \omega_i^2 k U(y_0) T - 3 \omega_r^2 k U(y_0) T \Big) \cos(k x_0-\omega_r t_0) \\
        &-e^{\omega_i(t_0+T)} \Big(\omega_i^2-\omega_r^2-k^2 U^2(y_0) + 2 \omega_r k U(y_0)\Big) \sin(k x_0-\omega_r (t_0+T)+k U(y_0) T) \\
        &-e^{\omega_i(t_0+T)} \Big(2 \omega_i \omega_r - 2 \omega_i k U(y_0)\Big) \cos(k x_0-\omega_r (t_0+T)+k U(y_0) T),
\end{split}
\end{equation}
\begin{equation}
\begin{split}
    f_5 &= -e^{\omega_i t_0}\omega_i \cos(k x_0-\omega_r t_0)+e^{\omega_i t_0}\Big(\omega_r-k U(y_0)\Big) \sin(k x_0-\omega_r t_0) \\
        &+ e^{\omega_i(t_0+T)}\omega_i \cos(k x_0-\omega_r (t_0+T)+k U(y_0) T) \\
        &- e^{\omega_i(t_0+T)}\Big(\omega_r-k U(y_0)\Big) \sin(k x_0-\omega_r (t_0+T)+k U(y_0) T),
\end{split}
\end{equation}
\begin{equation}
\begin{split}
    f_6 &= e^{\omega_i t_0}\omega_i \sin(k x_0-\omega_r t_0) +e^{\omega_i t_0}\Big(\omega_r-k U(y_0)\Big) \cos(k x_0-\omega_r t_0) \\
        &- e^{\omega_i(t_0+T)}\omega_i \sin(k x_0-\omega_r (t_0+T)+k U(y_0) T) \\
        &- e^{\omega_i(t_0+T)} \Big(\omega_r-k U(y_0)\Big) \cos(k x_0-\omega_r (t_0+T)+k U(y_0) T),
\end{split}
\end{equation}
\begin{equation}
\begin{split}
    g   &= \omega_i^2 + \omega_r^2 - 2 \omega_r k U(y_0) + k^2 U^2(y_0).
\end{split}
\end{equation}

The baseflow profile $U$ = $U(y_0)$ and the perturbation modes $\hat{u}(y_0)$ and $\hat{v}(y_0)$ are functions of the initial $y$-location of the tracers. 

With the flow map of the perturbed flow available, we can compute the deformation gradient and Cauchy-Green strain tensor and apply \eqref{eq:kappa} to calculate the Lagrangian curvature change. 
We determine the leading order terms of $\bar{\kappa}_{t_0}^{t_0+T}$, for the case of $\mathbf{r}'$ = $[1, \, 0]^{\top}$ and $\kappa_0$ = 0, by performing a Taylor series expansion in the perturbation $\epsilon$ and collecting the first-order terms.
To the first order, the curvature change for horizontal material lines in a perturbed parallel shear flow can be computed as 
\begin{equation}\label{eq:kappa_lot}
\begin{split}
    \bar{\kappa}_{t_0}^{t_0+T} &=
    \frac{k^2}{\omega_i^2 + \omega_r^2 - 2 \omega_r k U(y_0) + k^2 U^2(y_0)}\bigg[\\
        &e^{\omega_i t_0}\Big(
        \big(-\omega_i\hat{v}_r(y_0)+(\omega_r-kU(y_0))\hat{v}_i(y_0)\big)\cos(kx_0-\omega_r t_0) \\
        &+\big(\omega_i\hat{v}_i(y_0)+(\omega_r-kU(y_0))\hat{v}_r(y_0)\big)\sin(kx_0-\omega_r t_0)
        \Big) \\
        &+e^{\omega_i(t_0+T)}\Big(
        \big(\omega_i\hat{v}_r(y_0)-(\omega_r-kU(y_0))\hat{v}_i(y_0)\big)\cos(kx_0-\omega_r(t_0+T)+kU(y_0)T) \\
        &-\big(\omega_i\hat{v}_i(y_0)+(\omega_r-kU(y_0))\hat{v}_r(y_0)\big)\sin(kx_0-\omega_r(t_0+T)+kU(y_0)T)
        \Big)
    \bigg]\epsilon + \mathcal{O}(\epsilon^2).
\end{split}
\end{equation}
This completes the proof of Theorem \ref{sec:Thm1}.

\begin{figure}
    \makebox[\textwidth][c]{
	\includegraphics[width=0.4\textwidth]{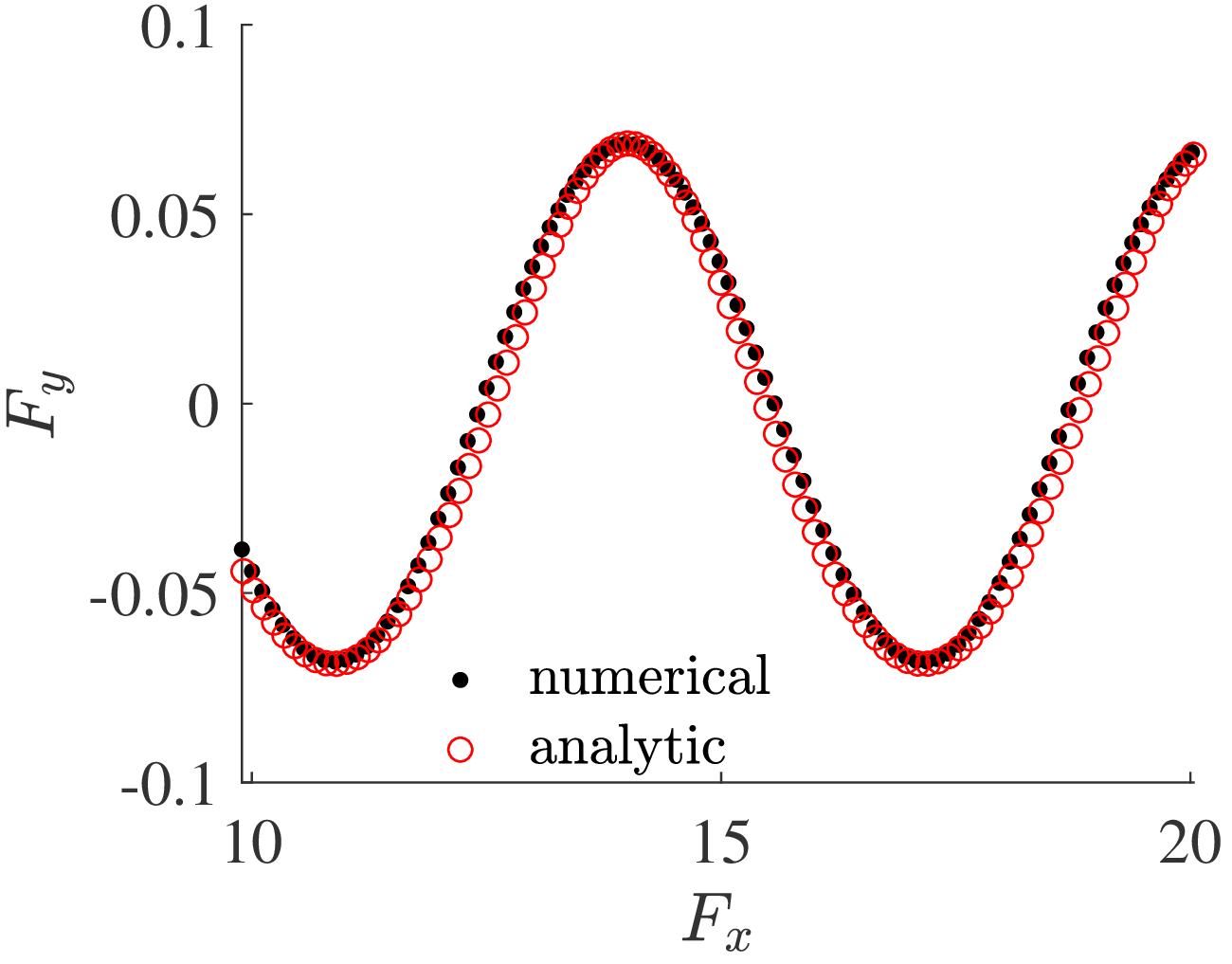}
	\hspace{0.1\textwidth}
	\includegraphics[width=0.4\textwidth]{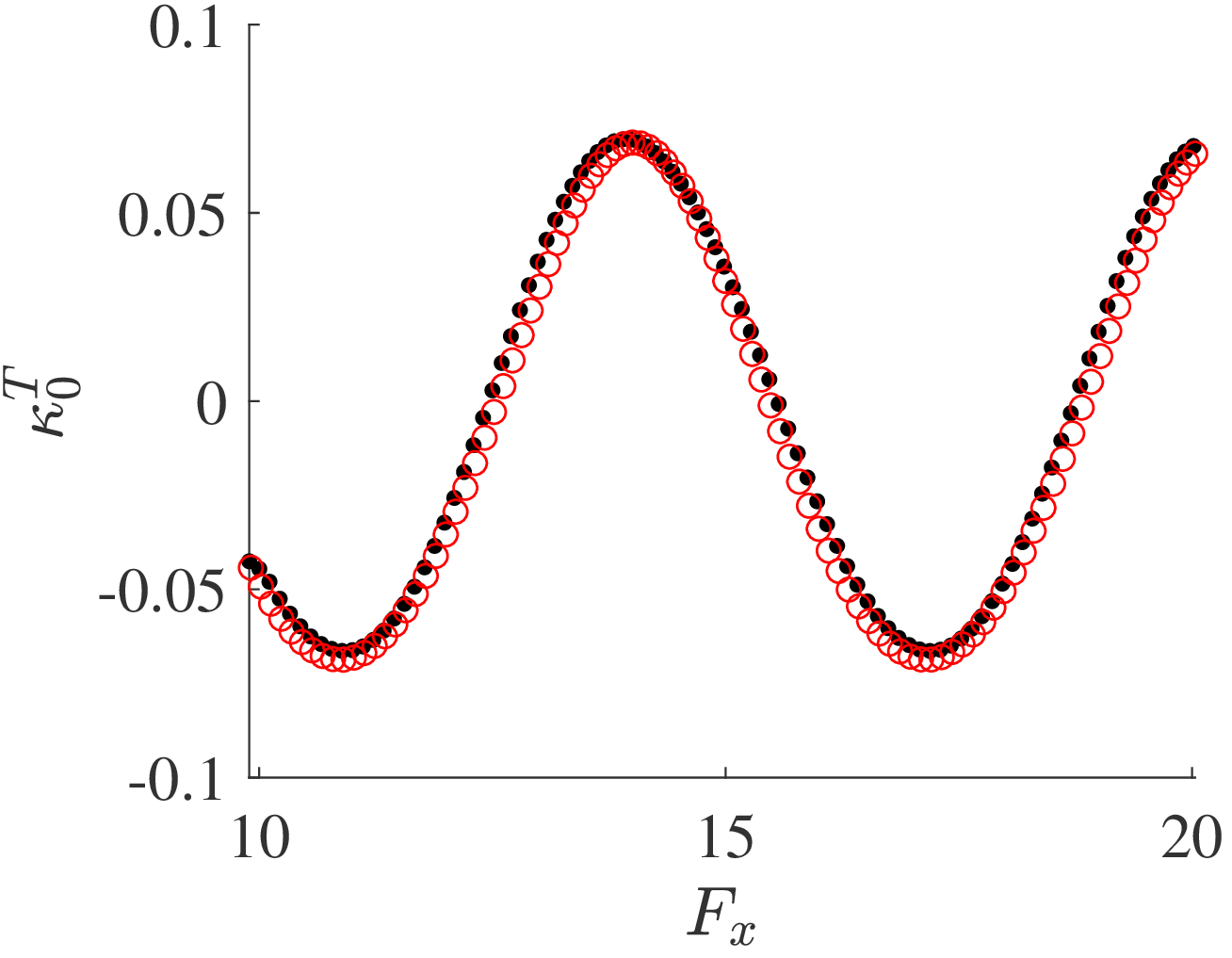}
	}
	\makebox[\textwidth][c]{
	\makebox[0.4\textwidth][c]{(a) Flow map}
    \hspace{0.1\textwidth}
    \makebox[0.4\textwidth][c]{(b) Curvature scalar}
    }
	\caption{Comparison of analytic and numerical approximations of the flow map $\mathbf{{{_\epsilon}F}}_{0}^{20}$ and the curvature scalar for $\bar{\kappa}_{0}^{20}$.
	Parameters: $\mathbf{u_0}=[(1+\tanh(y))/2, 0]^\top$, $y_0$ = 0, $k$ = 1, $\omega_r$ = 2, $\omega_i$ = 0.1, $\epsilon$ = 1\%, and $T$ = 20.}
	\label{fig:analytic_numeric}
\end{figure}

We check the analytic solutions of the flow map \eqref{eq:flowmap_eps} and the curvature scalar \eqref{eq:kappa_lot} for a material line advected under $\mathbf{u_0}=[(1+\tanh(y))/2, 0]^\top$ (hyperbolic tangent shear profile) and time-dependent perturbations of the form $\mathbf{u^\prime}$ = $\operatorname{Re}([\hat{u}(y),\hat{v}(y)]e^{i(kx - \omega t)})$.
The tracers are initialized at $y_0$ = 0 and $x_0$ $\in[0, 10]$ and the parameters are chosen to be $k$ = 1, $\omega_r$ = 2, $\omega_i$ = 0.1, $\epsilon$ = 1\%, and $T$ = 20. 
The eigenmode shape functions are set to $\hat{u}_r(y)$ = $\hat{u}_i(y)$ = $-\tanh(y)\sech(y)/k$ and $\hat{v}_r(y)$ = $-\hat{v}_i(y)$ = $\sech(y)$, such that a solenoidal velocity field is constructed.
In figure \ref{fig:analytic_numeric}, the analytic solutions from \eqref{eq:flowmap_eps} and \eqref{eq:kappa_lot} are compared to their numerical results from integrating $\mathbf{u_\epsilon=u_0+\epsilon u^\prime}$ described above. 
The deviations are small confirming the validity of our analytical approximation.

\begin{figure}
    \makebox[\textwidth][c]{
	\includegraphics[width=0.439\textwidth]{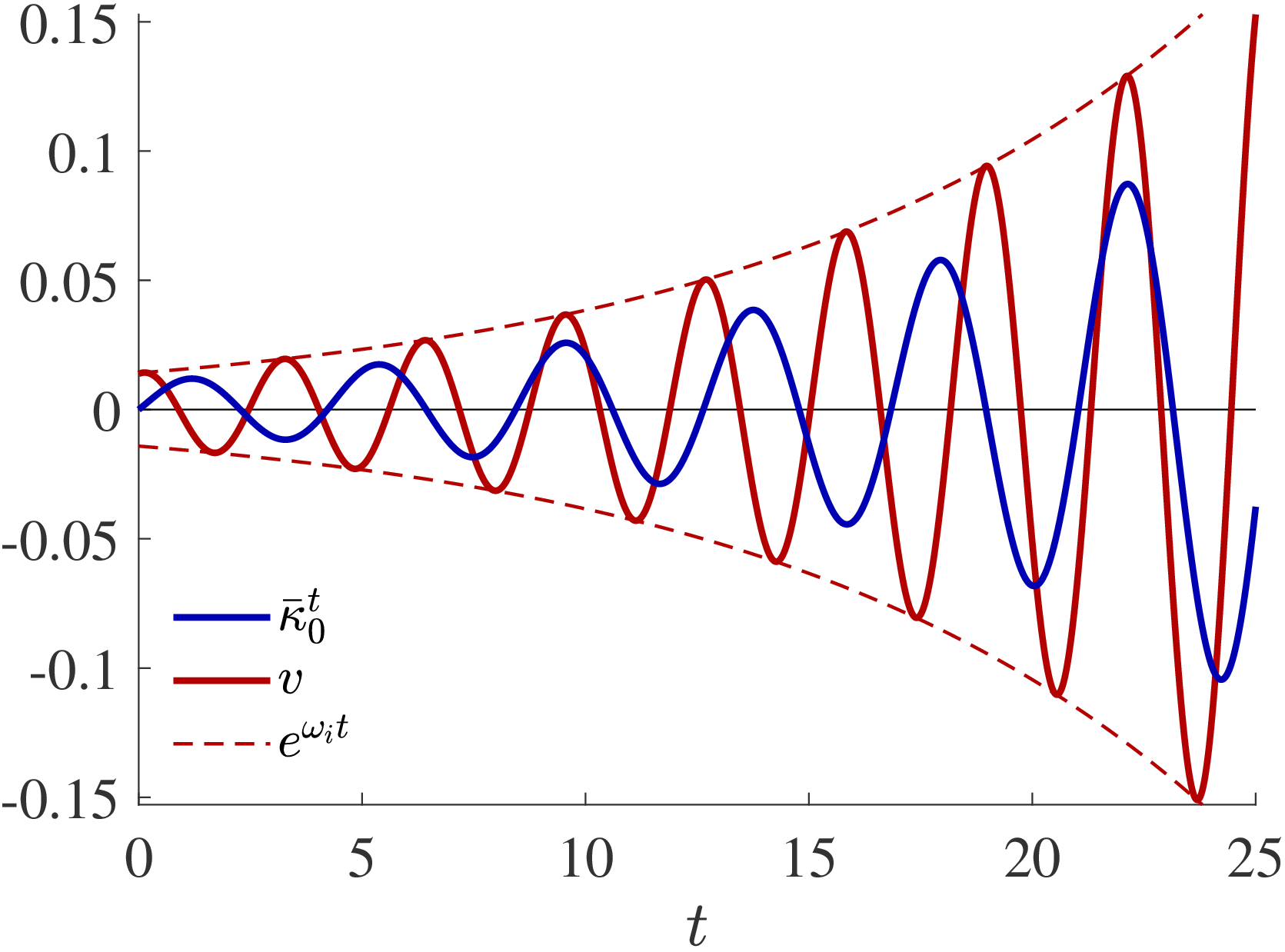}
	\hfill
	\includegraphics[width=0.45\textwidth]{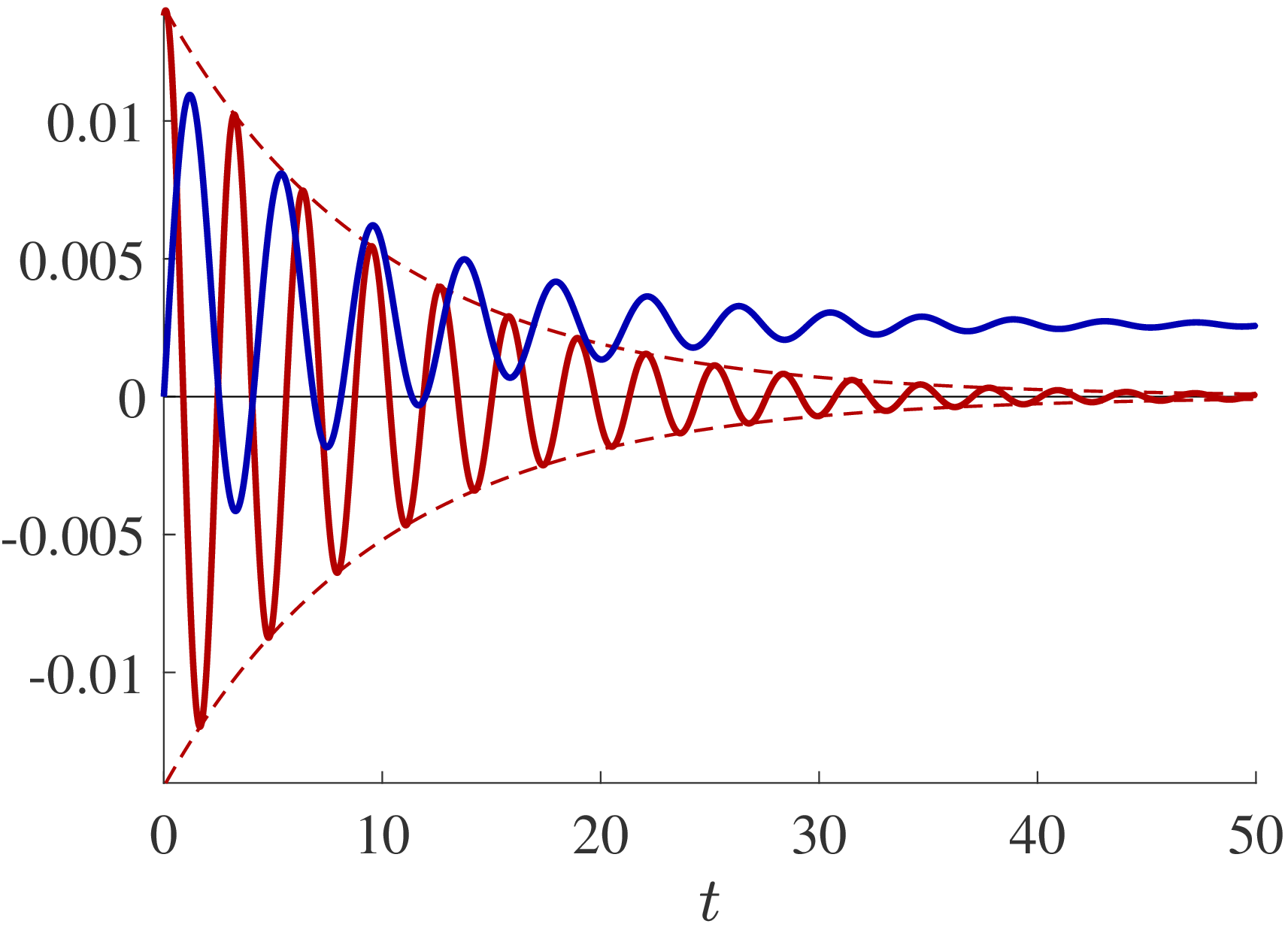}
	}
	\makebox[\textwidth][c]{
	\makebox[0.439\textwidth][c]{(a) $\omega_i$ > 0}
    \hfill
    \makebox[0.45\textwidth][c]{(b) $\omega_i$ < 0}
    }
	\caption{Comparison of analytic curvature scalar \eqref{eq:kappa_lot}, the velocity $v$ = $\operatorname{Re}(\hat{v}(y)e^{i(kx - \omega t)})$, and the envelope function $e^{\omega_i t}$.
	Parameters: $\mathbf{u_0}=[(1+\tanh(y))/2, 0]^\top$, $y_0$ = 0, $k$ = 1, $\omega_r$ = 2, $|\omega_i|$ = 0.1, $\epsilon$ = 1\%.}
	\label{fig:kt0TLOT}
\end{figure}

The temporal development of the curvature scalar \eqref{eq:kappa_lot} is illustrated in figure \ref{fig:kt0TLOT} together with the velocity $v$ = $\operatorname{Re}(\hat{v}(y)e^{i(kx - \omega t)})$ for positive (growing perturbation) and negative (ceasing perturbation) temporal growth rates.
The set of parameters is the same as introduced above, but for a single point $[x_0,y_0]$ = $[1,0]$ instead of a material line.

\section{Material response to a traveling wave} \label{appendix_wave}
\subsection{Model problem: traveling sine wave}

To understand the  kinematic response of fluid material
to  a traveling wave in general, the exponentially growing velocity field of a traveling sine wave is considered and as given by  $[u, v]^\top$ = $[0, Ae^{\omega_i t}\sin{(k c_r t - kx)}]$. 
The wave amplitude, speed and wave number are set to $A$ = 0.01, $c_r$ = 0.5 and $k$ = 4.2, respectively.

Integration of the kinematic
equation yields the particle coordinates $x_p$ = $\int_{t_0}^{t_1}u dt$ and $y_p$ = $\int_{t_0}^{t_1}v dt$. Because $u$ = 0 
there is no motion of the fluid material in $x$-direction. 
If the velocity amplitude is constant ($\omega_i$ = 0), then the $y$-location of a material line initialized at $y_p(t_0{=}0)$ = 0 is 
\begin{equation}
    y_p(x,t) = \frac{A}{k c_r}\left[\cos{(kx)}-\cos{(k c_r t - kx)}\right].
\label{eq:ypmode}
\end{equation}
Hence, the wave length of this traveling material wave  is the same
as that of the velocity wave. The former, however, travels at half the phase speed as compared to the latter. This can be   understand by inspecting the $x$-location of the particle
mode evaluated at $y_p$ = 0 from (\ref{eq:ypmode})
\begin{equation}
    \left. x\right\vert_{y_p{=}0} = \frac{1}{2}c_r t.
\end{equation}

For an exponentially increasing velocity field ($\omega_i$ = 0.9), the $y$-location of a material line initialized at $y_p(t_0{=}0)$ = 0 is 

\begin{align}
    y_p(x,t) =& \frac{A}{\omega_i^2+(k c_r)^2} \left[      \omega_i\sin{(kx)}+kc_r\cos{(kx)} + \right. \\
              & \left. e^{\omega_i   t}(\omega_i \sin{(kc_rt-kx)}-kc_r \cos{(kc_rt-kx)}) \right].
\end{align}
 Asymptotically for $t$ $\to$ $\infty$, the material  mode locks on to the velocity mode
 as is again understood by inspecting $x$ evaluated at $y_p$=0.
\begin{equation} \label{eq:phase_shift}
    \lim_{t\to\infty}\left. x\right\vert_{y_p{=}0} = c_r t-\frac{1}{k}\arctan\left(\frac{kc_r}{\omega_i}\right).
\end{equation}

\begin{figure}
    \makebox[\textwidth][c]{
	\includegraphics[height=0.25\textwidth]{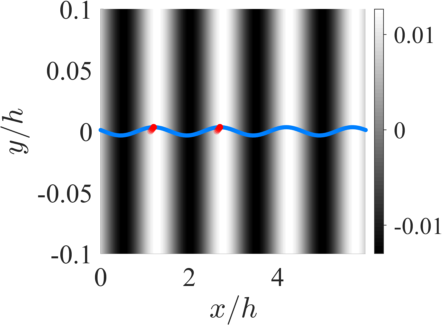}
	\hfill
	\includegraphics[height=0.25\textwidth]{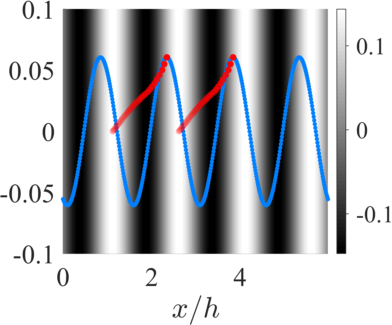}
	\hfill
	\includegraphics[height=0.25\textwidth]{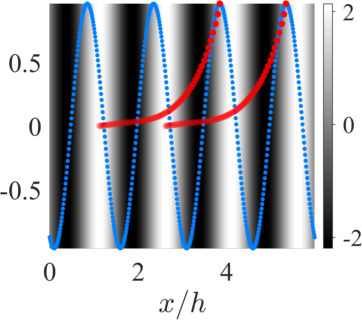}
	}
	\makebox[\textwidth][c]{
	\makebox[0.3\textwidth][c]{(a) $t$ = 0.3}
    \hfill
    \makebox[0.3\textwidth][c]{(b) $t$ = 3.0}
    \hfill
    \makebox[0.3\textwidth][c]{(c) $t$ = 6.0}
	}
	\caption{(a--c) Development of a material line (blue) under the velocity field $[u, v]^\top$ = $[0, Ae^{\omega_i t}\sin{(k c_r t - kx)}]$ (black-white) over time. Location history of Lagrangian curvature peaks in red.}
	\label{fig:sine}
\end{figure}

\begin{figure}
    \centering
	\includegraphics[height=0.27\textwidth]{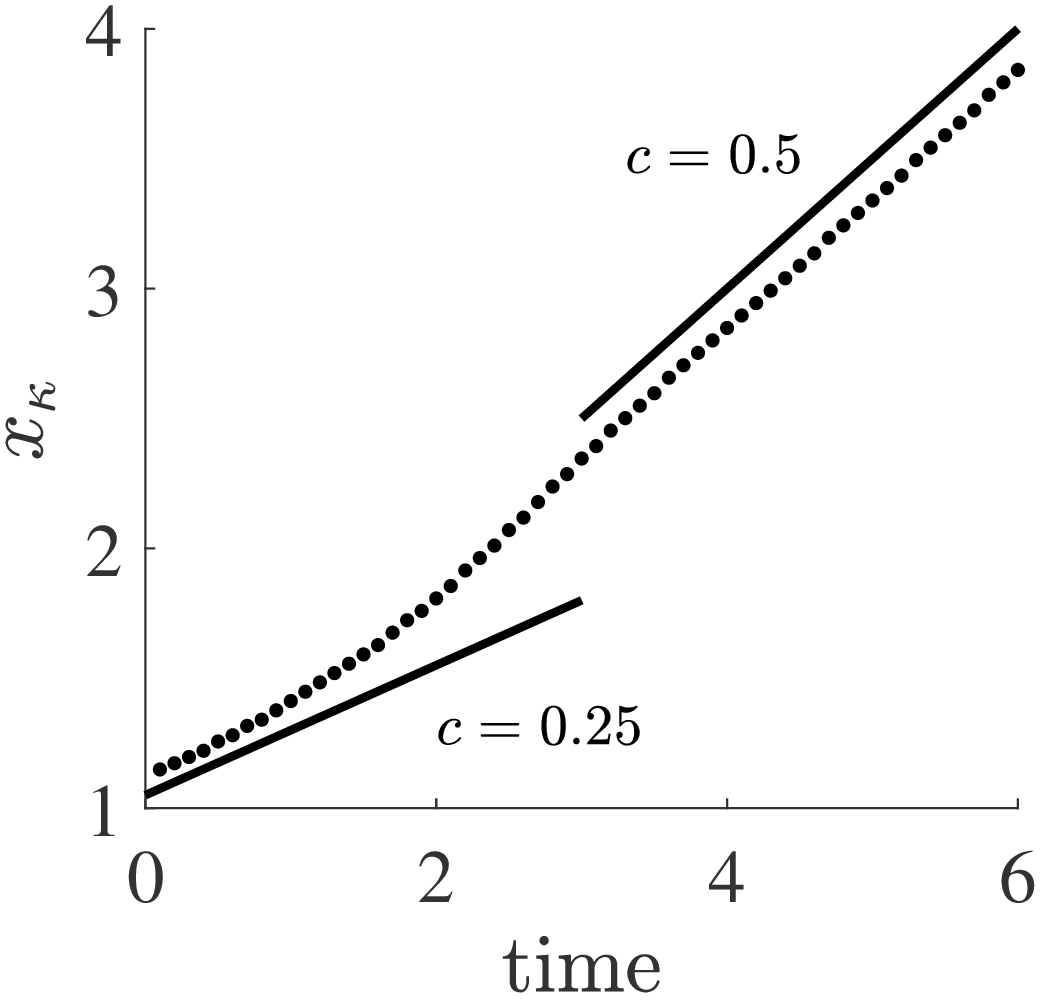}
	\caption{X-locations of the first Lagrangian curvature peak (see figure \ref{fig:sine}) over time. Phase velocity $c$ (slopes) indicated as solid lines in black.}
	\label{fig:peak_over_t}
\end{figure}

Figure \ref{fig:sine}(a--c) shows the development of a material line (blue) subject to an exponentially growing velocity mode (black-white), with the location history of curvature peaks in red. 
The $x$-location of the first curvature peak over time is shown in figure \ref{fig:peak_over_t}, where the slope confirms that the curvature mode initially travels at half the phase speed of the velocity mode but accelerates until they move uniformly.
The plots in (a) to (c) confirm the locking of the modes and shows that curvature and velocity peaks travel separated by the phase shift introduced in \eqref{eq:phase_shift}.
If wavenumber and phase speed are known, the growth rate $\omega_i$ can therefore be computed solely based on the shift between the curvature and the velocity mode.

\subsection{Traveling mode in the jet flow}

\begin{figure}
    \makebox[\textwidth][c]{
	\includegraphics[width=0.495\textwidth]{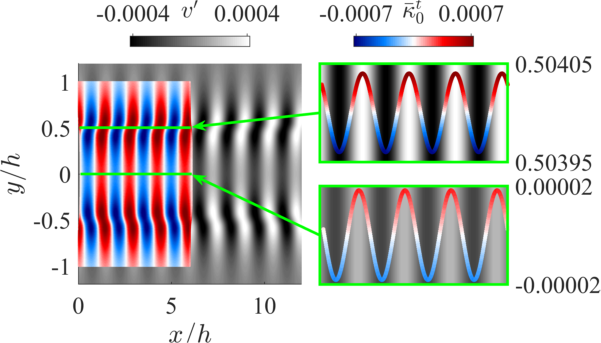}
	\hfill
	\includegraphics[width=0.485\textwidth]{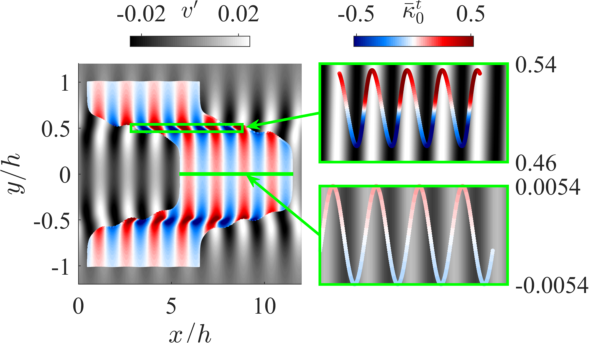}
	}
	\makebox[\textwidth][c]{
	\makebox[0.495\textwidth][c]{(a) $t$ = 0.1}
    \hfill
    \makebox[0.485\textwidth][c]{(b) $t$ = 5.0}
    }
	\caption{Curvature scalar field $\bar{\kappa}_{0}^{t}$ (blue-white-red) graphed over transverse velocity perturbation field $v'$ (black-white) at times $t$ = 0.1 (a) and $t$ = 5.0 (b). Initial perturbation with eigenmodes for wavenumber $k_{pert}$ = 4.2.
	(a) and (b) each have an overview plot on the left and detail plots of material lines along the shear layer and center line on the right. 
	Location of material lines and detail plots pointed out by green arrows and boxes. 
	Y-axis of detail plots is strongly stretched to show fluid particle motion (see green boxes). 
	Note that the color maps are adjusted at each time step.}
	\label{fig:vkappa}
\end{figure}

The shifting of velocity and curvature modes for the jet flow is visualized in figure \ref{fig:vkappa}. 
For a very short integration interval, the Lagrangian curvature field is approximately equal to $\dot{\kappa}_t$ and directly matches the pattern of the Eulerian transverse velocity.
While the material line motion initially corresponds to $v'$ (see figure \ref{fig:vkappa}a), the modes develop a phase shift in the jet core and the outer region as the velocity modes travel relative to the Lagrangian particle trace (see figure \ref{fig:vkappa}b).
Only along the shear layer, where the phase speed closely matches the advection velocity of the fluid, extrema of $\bar{\kappa}_{t_0}^{t_0+T}$ and $v'$ move together and result in much larger deformations of the particle trace than in the core flow (compare right plots of figure \ref{fig:vkappa}b).
From a Lagrangian perspective, the fluid particles in the shear layer see the perturbation mode as standing wave and therefore diverge faster than particles outside or inside the jet.

\bibliographystyle{unsrt}
\bibliography{bib}

\end{document}